\documentclass[pra,showpacs,twocolumn,aps]{revtex4-1}

\usepackage{amsmath}
\usepackage{amssymb}
\usepackage{latexsym}
\usepackage{amsfonts}
\usepackage{epsfig}
\usepackage{psfrag}
\usepackage{graphicx}

\usepackage{epsfig,color}
\usepackage{psfrag}
\usepackage[utf8]{inputenc}
\definecolor{darkgreen}{rgb}{0,0.4,0}

\newcommand{\ket}[1]{\left|#1\right>}
\newcommand{\bra}[1]{\left<#1\right|}   

\newcommand{\nn}{\nonumber\\}

\newcommand{\f}[1]{\mbox{\boldmath$#1$}}
\newcommand{\fk}[1]{\mbox{\boldmath$\scriptstyle#1$}}

\newcommand{\bea}{\begin{eqnarray}}
\newcommand{\ea}{\end{eqnarray}}
\newcommand{\eea}{\end{eqnarray}}
\newcommand{\ord}{\,{\cal O}}
\newcommand{\li}{\,\widehat{\cal L}}
\newcommand{\tr}{\,{\rm Tr}}

\begin{document}

\title
{
Equilibration and prethermalization in the Bose-Hubbard and Fermi-Hubbard models
}

\author{
F.~Queisser$^{1,2}$,
K.~V.~Krutitsky$^{1}$,
P.~Navez$^{1}$, 
and R.~Sch{\"u}tzhold$^{1,}$
}

\email{ralf.schuetzhold@uni-due.de}

\affiliation{
$^1$Fakult\"at f\"ur Physik, Universit\"at Duisburg-Essen, 
Lotharstrasse 1, 47057 Duisburg, Germany
\\
$^2$Department of Physics, University of British Columbia, Vancouver, 
V6T 1Z1 Canada
}

\date{\today}

\begin{abstract}
We study the Bose and Fermi Hubbard model in the (formal) limit of large 
coordination numbers $Z\gg1$.
Via an expansion into powers of $1/Z$, we establish a hierarchy of 
correlations which facilitates an approximate analytical derivation of the 
time-evolution of the reduced density matrices for one and two sites etc.
With this method, we study the quantum dynamics (starting in the ground state) 
after a quantum quench, i.e., after suddenly switching the tunneling rate 
$J$ from zero to a finite value, which is still in the Mott regime.  
We find that the reduced density matrices approach a (quasi) equilibrium 
state after some time.
For one lattice site, this state can be described by a thermal state
(within the accuracy of our approximation). 
However, the (quasi) equilibrium state of the reduced density matrices for 
two sites including the correlations cannot be described by a thermal state.
Thus, real thermalization (if it occurs) should take much longer time. 
This behavior has already been observed in other scenarios and is sometimes 
called ``pre-thermalization.'' 
Finally, we compare our results to numerical simulations for finite lattices 
in one and two dimensions and find qualitative agreement. 
\end{abstract}


\maketitle

\section{Introduction}
\label{section-Introduction}

Despite decades of research, our understanding of the quantum dynamics of 
interacting many-particle systems is still far from complete. 
One of the major unsolved questions (or rather a set of questions) 
is the problem of thermalization of isolated quantum
systems~\cite{D91,S94,BBS03,BBW04,CC06,CC07,RDYO07,RDO08,EHKKMWW09,MK09,CR10,PSSV11,KWE11,GME11,KISD11,BCH11,RS12,R13}. 
In one version, this question can be posed in the following way:
Given an interacting quantum many-body system on an infinite lattice in a 
globally excited state, do all observables involving a finite number of 
lattice sites settle down 
to a value which is consistent with a thermal state described by a suitable 
temperature?
Note that we do not consider thermalization induced by the coupling to some 
large thermal reservoir, but the intrinsic mechanism occurring in closed 
quantum systems during unitary evolution.

The global nature of the excitation is necessary because a local excitation 
(with a finite total energy) would typically disperse to infinity and leave 
the system locally at its ground state after some time. 
One option to create such a global excitation is a quantum quench:
Starting in the ground state of a given Hamiltonian, one suddenly 
(or at least non-adiabatically) changes some of the parameters,  
e.g., the external magnetic field or a coupling strength, and thus induces 
a global departure from the ground state (of the modified Hamiltonian).

This behavior crucially depends on the structure of the Hamiltonian.
Integrable models, for example, possess an infinite set of non-trivial 
conserved quantities.  
If these conserved quantities are measurable with local observables, 
there is no real thermalization.  
Instead, one should describe the state by a generalized Gibbs ensemble 
which contains a Lagrange multiplier for each conserved quantity.
This motivates the study of non-integrable models, such as the Bose-Hubbard 
model and Fermi-Hubbard model in more than one dimension considered here.
The models are prototypical examples for simple and yet non-trivial 
lattice Hamiltonians and can also be realized experimentally, for example,
with ultra-cold atoms in optical lattices~\cite{JZ07,MO06,LSADSS07,BDZ08,Y09,LSA12}.

Even if the thermalization occurs, there is still the question of 
the time scales involved, for example:
How fast does the system thermalize and do some observable thermalize faster 
than others? 
Are there intermediate stages and how fast do the quantum correlations spread?
The last question is related to the others since the unitary evolution of a
closed quantum system implies that an initially pure state will remain pure.
Hence the description of a local state by a thermal (i.e., mixed) density 
matrix is only possible due to quantum correlations with some remote part
of the lattice which is averaged over.

Quantum quenches have been considered before, for bosons and for fermions. 
For bosons, many studies have been devoted to one spatial dimension by 
employing exact diagonalization~\cite{KLA07,BKLO08,GR09,GR10},
time-dependent density matrix renormalization group theory
(t-DMRG)~\cite{KLA07,CDEO08,CFMSE08,FCMSE08,LK08,BRK11,BPBRK12}, 
and Jordan-Wigner fermionization~\cite{BPCK12}.
For corresponding experiments, see Refs.~\cite{GMHB02,KWS06,ST12,CBPE12,GKLKRSMADS12}. 
However, thermalization in one spatial dimension is quite different from 
the behavior in higher dimensions because quasi-particles in one dimension 
cannot thermalize via elastic two-body collisions due to energy-momentum 
conversation.

For bosons in higher dimensions, many of the methods which work well in one 
dimension cannot be applied.
Apart from some general statements concerning the relaxation of a quantum 
system towards equilibrium~\cite{CDEO08,CFMSE08,FCMSE08},
quantum quenches have been 
studied by using certain approximations, such as 
Bogoliubov-type approximations or strong-coupling perturbation 
theory~\cite{SUXF06,FSU08,NM13a,NM13b},
the Gutzwiller approximation~\cite{NHM11},
or related (semi) classical methods~\cite{SB08,SB10,SWD12},
as well as (truncated) exact diagonalization~\cite{KLA07}.
However, these approximations are only reliable in certain regions of
parameter space.
For an experimental realization of the quench from the Mott-insulator to the 
superfluid regime, see Ref.~\cite{CWBD11}.
 
For fermions in one spatial dimension, the integrability of the Fermi-Hubbard 
model facilitates the derivation of the exact evolution after a quench 
including effects such as ``pre-thermalization''~\cite{KE08,KWE11,HU13}.  
Again, in higher dimensions, appropriate approximations are necessary,
such as a time-dependent Monte-Carlo method~\cite{GA12},
time-dependent dynamical mean field theory~\cite{EKW09,EKW10,A10,A11,WTE12},
the Gutzwiller ansatz for fermions~\cite{SF10,SF11,SSF12},
the flow equation  method~\cite{MK08,MK09,MK10,EHKKMWW09},
or effective quasi-particle methods~\cite{SGJ10}.

In the present work, we study the quantum evolution after a quench in the 
Bose and Fermi Hubbard models.
We develop and employ an analytic approximation technique which is 
controlled by an expansion into powers of the inverse coordination 
number $1/Z$ (see also~\cite{HSH09}).
Note that the $1/Z$-expansion employed here is somewhat similar to 
time-dependent dynamical mean field theory (t-DMFT), but the $1/\sqrt{Z}$ 
scaling of the hopping term in the Hamiltonian (used in t-DMFT) is replaced 
by a $1/Z$ scaling in our approach -- which allows us to derive analytic 
expressions for the time-dependent correlation functions after the quench. 

\section{Bose-Hubbard Model}\label{section-Bose-Hubbard-Model}

The Bose-Hubbard model is one of the most simple and yet non-trivial models 
in condensed matter theory~\cite{FWGF89,J98,Z03}.
It describes identical bosons hopping on a lattice with the tunneling 
rate $J$.
In addition, two (or more) bosons at the same lattice site repel each other 
with the interaction energy $U$.
The Hamiltonian reads 
\bea
\label{Bose-Hubbard-Hamiltonian}
\hat H
=
-\frac{J}{Z}\sum_{\mu\nu} T_{\mu\nu} \hat b^\dagger_\mu \hat b_\nu
+\frac{U}{2}\sum_{\mu} 
\hat n_{\mu}(\hat n_{\mu}-1)
\,.
\ea
Here $\hat b^\dagger_\mu$ and $\hat b_\nu$ are the creation and annihilation 
operators at the lattice sites $\mu$ and $\nu$, respectively, which obey 
the usual commutation relations 
\bea
\left[\hat b_\nu,\hat b_\mu^\dagger\right]=\delta_{\mu\nu}
\;,\,
\left[\hat b_\nu^\dagger,\hat b_\mu^\dagger\right]=
\left[\hat b_\nu,\hat b_\mu\right]=0
\,.
\ea
The lattice structure is encoded in the adjacency matrix $T_{\mu\nu}$ 
which equals unity if $\mu$ and $\nu$ are tunneling neighbors 
(i.e., if a particle can hop from $\mu$ to $\nu$) and zero otherwise. 
The number of tunneling neighbors at a given site $\mu$ yields the 
coordination number $Z=\sum_\nu T_{\mu\nu}$
(we assume a translationally invariant lattice). 
Finally, $\hat n_{\mu}=\hat b^\dagger_\mu \hat b_\mu$ is the number 
operator and we assume unit filling $\langle\hat n_{\mu}\rangle=1$ 
in the following. 
Note that the total particle number 
$\hat N=\sum_\mu \hat n_\mu$ is conserved $[\hat H,\hat N]=0$. 

The Bose-Hubbard model is considered as one of the prototypical examples for 
a quantum phase transition~\cite{sachdev}.
If the interaction term dominates $U\gg J$, the bosons are pinned to 
their lattice sites and we have the Mott insulator state 
\bea
\label{Mott-state}
\ket{\Psi_{\rm Mott}^{J=0}}
=
\bigotimes\limits_\mu\ket{1}_\mu
=
\prod\limits_\mu\hat b_\mu^\dagger\ket{0}
\;\leadsto\;
\hat H\ket{\Psi_{\rm Mott}^{J=0}}=0
\,,
\ea
which is fully localized. 
If the hopping rate dominates $U\ll J$, on the other hand, the particles 
can propagate freely across the lattice and become completely delocalized 
\bea
\label{superfluid-state}
\ket{\Psi_{\rm superfluid}^{U=0}}
&=&
\frac{1}{\sqrt{N! N^N}}
\left(\sum_{\mu}\hat b_\mu^\dagger\right)^N\ket{0}
\nn
&=&
\frac{1}{\sqrt{N!}}
\left(\hat b_{\fk{k}=0}^\dagger\right)^N\ket{0}
\,,
\ea
which is the superfluid phase. 
Obviously, the Mott state~(\ref{Mott-state}) does not have any correlations
\cite{footnote-correlations} between lattice sites, for example 
$\langle\hat b^\dagger_\mu \hat b_\nu\rangle_{\rm Mott}=\delta_{\mu\nu}$, 
whereas the superfluid state in~(\ref{superfluid-state}) shows correlations 
across the whole lattice 
$\langle\hat b^\dagger_\mu \hat b_\nu\rangle_{\rm superfluid}=1$. 
Furthermore, the Mott insulator state is separated by a finite energy gap 
from the lowest excited state, while the superfluid state possesses 
sound-like modes with arbitrarily low energies 
(for an infinitely large lattice $N\to\infty$). 
Finally, the Bose-Hubbard model can be realized experimentally 
(to a very good approximation) with ultra-cold atoms in optical 
lattices~\cite{B05,S07,RSN97}
and it was even possible to observe the aforementioned
phase transition in these systems~\cite{G02}.  

In spite of its simplicity, the Bose-Hubbard 
model~(\ref{Bose-Hubbard-Hamiltonian}) cannot be solved analytically. 
Numerical simulations are limited to reduced sub-spaces or 
small systems sizes, see Section~\ref{Numerical} below.
Analytical approaches are based on suitable approximations.
In order to control the error of these approximations, they should be 
based on an expansion in term of some large or small control parameter. 
For the Bose-Hubbard model~(\ref{Bose-Hubbard-Hamiltonian}), one could 
consider the limit of large $\langle\hat n_{\mu}\rangle\gg1$  
or small $\langle\hat n_{\mu}\rangle\ll1$ filling~\cite{SUXF06,FSU08},
for example, or the limit of weak coupling $U\ll J$ or strong coupling 
$U\gg J$~\cite{FM94,FM96,DZ06,FKKKT2009}.
However, none of these limits is particularly well suited for studying 
the Mott--superfluid phase transition. 
To this end, we consider the limit $Z\gg1$ in the following and employ 
an expansion into powers of $1/Z$ as small control parameter. 
Note that an expansion in powers of $1/Z$ was also used to derive bosonic 
dynamical mean-field equations (which were then solved numerically) 
in~\cite{HSH09,LBHH11,LBHH12}.

\section{Hierarchy of Correlations}
\label{hierarchyofcorr}

Let us consider general Hamiltonians of the form 
\bea
\hat H=\frac1Z\sum_{\mu\nu}\hat H_{\mu \nu}+\sum_\mu\hat H_\mu
\,,
\ea
which includes the bosonic and fermionic 
Hubbard models~(\ref{Bose-Hubbard-Hamiltonian}) 
and~(\ref{Fermi-Hubbard-Hamiltonian}) as special cases.
The quantum evolution of the density operator $\hat\rho$ describing the 
state of the full lattice can be written as 
\bea
\label{Liouville}
i\partial_t\hat\rho 
= \left[\hat H,\hat\rho\right] 
&=&
\frac1Z\sum_{\mu\nu}\left[\hat H_{\mu \nu},\hat\rho\right] 
+\sum_\mu\left[\hat H_\mu,\hat\rho\right] 
\nn
&=& 
\frac1Z\sum_{\mu\nu}\li_{\mu \nu}\hat\rho 
+
\sum_\mu\li_\mu\hat\rho
\,,
\ea
where we have introduced the Liouville super-operators $\li_{\mu \nu}$
and $\li_\mu$ as short-hand notation.
As the next step, we introduce the reduced density matrices for one 
or more lattice sites via averaging (tracing) over all other sites
\bea
\label{reduced-density-matrices}
\hat\rho_\mu 
&=&
\tr_{\not\mu}\{\hat\rho\}
\;,\quad
\hat\rho_{\mu\nu}
=
\tr_{\not\mu\not\nu}\{\hat\rho\}
\,,
\ea
and so on. 
Note that $\tr\{\hat\rho\}=1$ implies $\tr_\mu\{\hat\rho_\mu\}=1$
and $\tr_{\mu\nu}\{\hat\rho_{\mu\nu}\}=1$ etc. 
Next we define correlated parts of the reduced density matrices via
\bea
\label{correlated-parts}
\hat\rho_{\mu\nu}
&=&
\hat\rho_{\mu\nu}^{\rm corr}+\hat\rho_{\mu}\hat\rho_{\nu}
\\
\hat\rho_{\mu\nu\lambda}
&=&
\hat\rho_{\mu\nu\lambda}^{\rm corr}+
\hat\rho_{\mu\nu}^{\rm corr}\hat\rho_{\lambda}+
\hat\rho_{\mu\lambda}^{\rm corr}\hat\rho_{\nu}+
\hat\rho_{\nu\lambda}^{\rm corr}\hat\rho_{\mu}+
\hat\rho_{\mu}\hat\rho_{\nu}\hat\rho_{\lambda}
\nonumber
\,,
\ea
and analogously for more lattice sites. 
As a consequence, we obtain from Eq.~(\ref{Liouville}) the evolution 
equation for the one-point density matrix
\bea
\label{one-site}
i\partial_t\hat\rho_{\mu}
=
\frac{1}{Z}
\sum_{\kappa\neq\mu}\tr_{\kappa}\left\{
\li^S_{\mu \kappa}
(\hat\rho^{\rm corr}_{\mu \kappa}+\hat\rho_\mu \hat\rho_\kappa)\right\}
+
\li_\mu\hat\rho_{\mu}
\,,
\eea
where $\li_{\mu \nu}^S=\li_{\mu \nu}+\li_{\nu \mu}$ denotes the 
symmetrized form. 
Obviously, solving this equation exactly requires knowledge of the 
two-point correlation $\hat\rho^{\rm corr}_{\mu \kappa}$.
The time-evolution of this quantity can also be obtained from 
Eq.~(\ref{Liouville}) and reads
\bea
\label{two-sites}
i \partial_t \hat\rho^{\rm corr}_{\mu \nu}
&=&
\li_\mu\hat\rho^{\rm corr}_{\mu\nu}
+
\frac1Z\li_{\mu\nu}
(\hat\rho^{\rm corr}_{\mu\nu}+\hat\rho_\mu\hat\rho_\nu)
\nn
&&
-
\frac{\hat\rho_{\mu}}{Z}
\tr_{\mu}
\left\{\li^S_{\mu\nu}
(\hat\rho^{\rm corr}_{\mu\nu}+\hat\rho_\mu\hat\rho_\nu)
\right\}
\nn
&&+
\frac1Z
\sum_{\kappa\not=\mu,\nu} 
\tr_{\kappa}
\left\{
\li^S_{\mu \kappa}
(\hat\rho^{\rm corr}_{\mu\nu\kappa}+
\hat\rho^{\rm corr}_{\mu\nu}\hat\rho_{\kappa}+
\hat\rho^{\rm corr}_{\nu\kappa}\hat\rho_{\mu})
\right\}
\nn
&&
+(\mu\leftrightarrow\nu)
\,.
\eea
As one would expect, this equation contains the three-point 
correlator $\hat\rho^{\rm corr}_{\mu\nu\kappa}$, and similarly 
the evolution equation for $\hat\rho^{\rm corr}_{\mu\nu\kappa}$
contains the four-point correlator etc.
In general, one cannot exactly solve this infinite set of equations.
However, the limit $Z\gg1$ facilitates an approximate solution
that can be systematically improved.

Let us start from an initial state
$\hat\rho^{\rm in}=\bigotimes_\mu\hat\rho^{\rm in}_\mu$
that does not have any correlations 
(i.e., $\hat\rho^{\rm corr}_{\mu \nu}(0)=0$ and 
$\hat\rho^{\rm corr}_{\mu\nu\kappa}(0)=0$, etc.)
such as the Mott state~(\ref{Mott-state}).
In this case, the right-hand side of Eq.~(\ref{two-sites})
scales as $1/Z$ and thus the time evolution creates only small correlations 
$\hat\rho^{\rm corr}_{\mu \nu}(t)$. 
If these correlations are small initially, 
$\hat\rho^{\rm corr}_{\mu \nu}(0)=\ord(1/Z)$,
they remain small at least for a finite time. 
The order of terms in the second line of Eq.~(\ref{two-sites}) is determined
by the correlated parts of the density matrices.
This is because the summation over $\kappa$ gives at most a factor of $Z$
which is compensated by the factor $1/Z$ in front of the sum.
In addition, we can neglect the term in Eq.~(\ref{one-site})
which contains $\hat\rho^{\rm corr}_{\mu \nu}$ because it is of the
higher order than the others. Thus, we arrive at 
an approximate equation containing one-point density matrices only 
\bea
\label{one-site-approx}
i\partial_t\hat\rho_{\mu}
=
\frac{1}{Z}
\sum_{\kappa\neq\mu}
\tr_{\kappa}
\left\{
\li^S_{\mu \kappa}
\hat\rho_\mu \hat\rho_\kappa
\right\}
+
\li_\mu\hat\rho_{\mu}
+
\ord(1/Z)
\;.
\eea
The approximate solution $\hat\rho_\mu^0$ of this self-consistent equation 
is valid to lowest order in $1/Z$, i.e., 
$\hat\rho_\mu=\hat\rho_\mu^0+\ord(1/Z)$ 
and reproduces the well-known Gutzwiller ansatz~\cite{G63,RK91,J98}.
If we now insert this approximate solution $\hat\rho_\mu^0$ into
Eq.~(\ref{two-sites}), we get an approximate evolution equation for 
the two-point correlator
\bea
\label{two-sites-approx}
i \partial_t \hat\rho^{\rm corr}_{\mu \nu}
&=&
\li_\mu\hat\rho^{\rm corr}_{\mu\nu}
+
\frac1Z\li_{\mu\nu}\hat\rho^0_\mu\hat\rho^0_\nu
-
\frac{\hat\rho^0_{\mu}}{Z}
\tr_{\mu}
\left\{\li^S_{\mu\nu}\hat\rho^0_\mu\hat\rho^0_\nu\right\}
\nn
&&
+
\frac1Z
\sum_{\kappa\not=\mu,\nu} 
\tr_{\kappa}
\left\{
\li^S_{\mu \kappa}
(\hat\rho^{\rm corr}_{\mu\nu}\hat\rho^0_{\kappa}+
\hat\rho^{\rm corr}_{\nu\kappa}\hat\rho^0_{\mu})
\right\}
\nn
&&
+(\mu\leftrightarrow\nu)
+\ord(1/Z^2)
\,.
\eea
Since we assumed that the three-point correlations 
$\hat\rho^{\rm corr}_{\mu\nu\kappa}$ are suppressed by $\ord(1/Z^2)$,
they do not spoil this line of arguments.
In complete analogy, it is possible to derive the 
evolution equations for any $\ell$-point function, 
see Appendix~\ref{hierarchyApp}.
Thus, we find that $\ell$-point correlations are suppressed as 
$\ord(1/Z^{\ell-1})$, i.e.,
\bea
\label{hierarchy}
\hat\rho_{\mu}
&=&
\ord\left(Z^0\right)
\;,
\quad
\hat\rho^{\rm corr}_{\mu\nu}
=
\ord\left(1/Z\right)
\;,
\nn
\hat\rho^{\rm corr}_{\mu\nu\kappa} &=& \ord\left(1/Z^2\right)
\;,
\quad
\hat\rho^{\rm corr}_{\mu\nu\kappa\lambda}
=
\ord\left(1/Z^3\right)
\;,
\ea
and so on, see Appendix~\ref{hierarchyApp}.
The hierarchy~(\ref{hierarchy}) is related to 
the quantum de~Finetti theorem~\cite{CKMR07}, 
the generalized cumulant expansion~\cite{K62}, 
and the Bogoliubov-Born-Green-Kirkwood-Yvon (BBGKY) hierarchy~\cite{B75},
but we are considering lattice sites instead of particles.
As an example for the four-point correlator, let us consider observables 
$\hat A_\mu$, $\hat B_\nu$, $\hat C_\kappa$, and $\hat D_\lambda$ at 
four different lattice sites, which have vanishing on-site expectation values 
$\langle\hat A_\mu\rangle=
\langle\hat B_\nu\rangle=
\langle\hat C_\kappa\rangle=
\langle\hat D_\lambda\rangle=0$. 
In this case, the hierarchy~(\ref{hierarchy}) implies 
\bea
\langle\hat A_\mu\hat B_\nu\hat C_\kappa\hat D_\lambda\rangle 
&=&
\langle\hat A_\mu\hat B_\nu\rangle 
\langle\hat C_\kappa\hat D_\lambda\rangle 
+
\langle\hat A_\mu\hat C_\kappa\rangle 
\langle\hat B_\nu\hat D_\lambda\rangle 
\nn
&&
+
\langle\hat A_\mu\hat D_\lambda\rangle 
\langle\hat B_\nu\hat C_\kappa\rangle 
+
\ord\left(1/Z^3\right)
\,,
\ea
which resembles the Wick theorem in free quantum field theory
(even though the quantum system considered here is strongly interacting). 

\section{Mott Insulator State}\label{mottsection}

Now let us apply the hierarchy discussed above to the Bose-Hubbard 
model~(\ref{Bose-Hubbard-Hamiltonian}).
To this end, we start with the factorizing Mott state~(\ref{Mott-state})
at zero hopping rate $J=0$ as our initial state
\bea
\label{init-state}
\hat\rho^{\rm in}
=
\bigotimes\limits_\mu\hat\rho^{\rm in}_\mu
=
\bigotimes\limits_\mu\ket{1}_\mu\!\bra{1}
\,.
\ea
Then we slowly switch on the hopping rate $J(t)$ until we reach its 
final value. 
In view of the finite energy gap, the adiabatic theorem implies that 
we stay very close to the real ground state of the system if we do 
this slowly enough. 
Of course, we cannot cross the phase transition in this way 
(i.e., adiabatically) since the energy gap vanishes at the critical point, 
see Section~\ref{Quench dynamics} below. 

Since we have $\langle\hat b_\mu\rangle=0$ in the Mott state, 
Eq.~(\ref{one-site-approx}) simplifies to 
\bea
\label{one-site-Mott}
i\partial_t\hat\rho_{\mu}
&\approx&
\frac{1}{Z}
\sum_{\kappa\neq\mu}\tr_{\kappa}\left\{
\li^S_{\mu \kappa}
\hat\rho_\mu \hat\rho_\kappa\right\}
+
\li_\mu\hat\rho_{\mu}
=0 
\nn
&&
\leadsto\,
\hat\rho_{\mu}^0=\ket{1}_\mu\!\bra{1}
\,.
\eea
Thus, to zeroth order in $1/Z$ (i.e., on the Gutzwiller mean-field level),
the Mott insulator state $\hat\rho_{\mu}^0$ for finite $J$ has the same 
form as for $J=0$.
To obtain the first order in $1/Z$, we insert this result 
into~(\ref{two-sites-approx}). 
Again using $\langle\hat b_\mu\rangle=0$, we find 
\bea
i \partial_t \hat\rho^{\rm corr}_{\mu \nu}
&=&
\left(\li_\mu+\li_\nu\right)\hat\rho^{\rm corr}_{\mu\nu}
+
\frac1Z\li_{\mu\nu}^S\hat\rho_\mu^0\hat\rho_\nu^0
\nn
&&+
\frac1Z
\sum_{\kappa\not=\mu,\nu} 
\tr_{\kappa}
\left\{
\li^S_{\mu\kappa}\hat\rho^{\rm corr}_{\nu\kappa}\hat\rho_{\mu}^0
+\li^S_{\nu\kappa}\hat\rho^{\rm corr}_{\mu\kappa}\hat\rho_{\nu}^0
\right\}
\nn
&&
+\ord(1/Z^2)
\,.
\eea
Formally, this is an evolution equation for an infinite dimensional 
matrix $\hat\rho^{\rm corr}_{\mu \nu}$.
Fortunately, however, it suffices to consider a few elements only. 
If we introduce 
$\hat p_\mu=\ket{1}_\mu\!\bra{2}$ 
and 
$\hat h_\mu=\ket{0}_\mu\!\bra{1}$ 
as local particle and hole operators
(these excitations are sometimes~\cite{KJSW90,CBPE12,BPCK12} 
called doublons and holons), all the interesting physics can 
be captured by their correlation functions (for $\mu\neq\nu$)
\begin{eqnarray}
f_{\mu\nu}^{11}
&=&
\langle\hat{h}^\dagger_\mu\hat{h}_\nu^{} \rangle 
=
\tr\left\{\hat{\rho}\,\hat{h}^\dagger_\mu\hat{h}_\nu^{}\right\}
=
\tr_{\mu\nu}\left\{
\hat{\rho}^{\rm corr}_{\mu\nu}\hat{h}^\dagger_\mu\hat{h}_\nu^{}\right\}
\,,
\nn
f_{\mu\nu}^{12}
&=&
\langle\hat{h}^\dagger_\mu\hat{p}_\nu^{} \rangle 
=
\tr\left\{\hat{\rho}\,\hat{h}^\dagger_\mu\hat{p}_\nu^{}\right\}
=
\tr_{\mu\nu}\left\{
\hat\rho^{\rm corr}_{\mu\nu}\hat{h}^\dagger_\mu\hat{p}_\nu^{}\right\}
\,,
\nn
f_{\mu\nu}^{21}
&=&
\langle\hat{p}^\dagger_\mu\hat{h}_\nu^{} \rangle 
=
\tr\left\{\hat{\rho}\,\hat{p}^\dagger_\mu\hat{h}_\nu^{}\right\}
=
\tr_{\mu\nu}\left\{
\hat{\rho}^{\rm corr}_{\mu\nu}\hat{p}^\dagger_\mu\hat{h}_\nu^{}\right\}
\,,
\nn
f_{\mu\nu}^{22}
&=&
\langle\hat{p}^\dagger_\mu\hat{p}_\nu^{} \rangle 
=
\tr\left\{\hat{\rho}\,\hat{p}^\dagger_\mu\hat{p}_\nu^{}\right\}
=
\tr_{\mu\nu}\left\{
\hat{\rho}^{\rm corr}_{\mu\nu}\hat{p}^\dagger_\mu\hat{p}_\nu^{}\right\}
\,.
\end{eqnarray}
To first order in $1/Z$, these correlation functions form a closed set of 
equations~\cite{QNS12}
\begin{eqnarray}
\label{f12-Mott}
i\partial_t f^{12}_{\mu\nu}
&=&
-\frac{J}{Z}\sum_{\kappa\neq \mu,\nu}
\left(T_{\mu\kappa}(f^{12}_{\kappa\nu}+\sqrt{2}f^{22}_{\kappa\nu})
\right.
\nn
&&
+
\left.
\sqrt{2}T_{\nu\kappa}(f^{11}_{\mu\kappa}+\sqrt{2}f_{\mu\kappa}^{12})\right)
\nn
&&
+
Uf^{12}_{\mu\nu}-\frac{J\sqrt{2}}{Z}T_{\mu\nu}
\;,
\end{eqnarray}
and
\begin{eqnarray}
i\partial_t f^{21}_{\mu\nu}
&=&
+\frac{J}{Z}\sum_{\kappa\neq \mu,\nu}
\left(T_{\nu\kappa}(f^{21}_{\kappa\mu}+\sqrt{2}f^{11}_{\kappa\mu})
\right.
\nn
&&
+
\left.
\sqrt{2}T_{\mu\kappa}(f^{22}_{\kappa\nu}+\sqrt{2}f_{\kappa\nu}^{12})\right)
\nn
&&
-
Uf^{21}_{\mu\nu}+\frac{J\sqrt{2}}{Z}T_{\mu\nu}
\;,
\end{eqnarray}
as well as 
\begin{eqnarray}
\label{f11-Mott}
i\partial_t f^{11}_{\mu\nu}
&=&
i\partial_t f^{22}_{\mu\nu}
=
-\frac{\sqrt{2}J}{Z}\sum_{\kappa\neq\mu,\nu}\left(T_{\mu\kappa}
f_{\kappa\nu}^{21}-T_{\nu\kappa}f_{\mu\kappa}^{12}\right)
\,.
\nn
\end{eqnarray}
This truncation is due to the fact that the correlation functions 
$f^{mn}_{\mu\nu}$ involving higher occupation numbers $m\geq3$ or $n\geq3$
do not have any source terms of order $1/Z$ and hence do not contribute
at that level. 
Exploiting translational symmetry, we may simplify these equations by a 
spatial Fourier transformation with 
\begin{eqnarray}
\label{T_k}
T_{\mu\nu}
&=&
\frac{Z}{N}\sum_\mathbf{k}T_{\mathbf{k}}
e^{i\mathbf{k}\cdot(\mathbf{x}_\mu-\mathbf{x}_\nu)}
\;,
\\
\label{f_k}
f^{ab}_{\mu\nu}
&=&
\frac{1}{N}\sum_\mathbf{k}f^{ab}_{\mathbf{k}}
e^{i\mathbf{k}\cdot(\mathbf{x}_\mu-\mathbf{x}_\nu)}\,,
\end{eqnarray}
where $N$ denotes the number of lattice sites 
(which equals the number of particles in our case).
Formally, in order to Fourier transform 
equations~(\ref{f12-Mott})-(\ref{f11-Mott}), one should add the summands  
corresponding to $\kappa=\mu$ and $\kappa=\nu$.
Since these terms are of order $1/Z^2$, they do not spoil our first-order 
analysis.  
However, when going to second order $1/Z^2$, (see Section~\ref{Z2} below), 
they have to be taken into account.

After the Fourier transformation~(\ref{T_k}) and~(\ref{f_k}),
Eqs.~(\ref{f12-Mott})-(\ref{f11-Mott}) become 
\begin{eqnarray}
(i\partial_t-U+3 J T_\mathbf{k})f_\mathbf{k}^{12}
&=&
-\sqrt{2}J T_\mathbf{k}(f_\mathbf{k}^{11}+f_\mathbf{k}^{22}+1)
\,,\label{diff0}
\\
(i\partial_t+U-3 J T_\mathbf{k})f_\mathbf{k}^{21}
&=&
+\sqrt{2}J T_\mathbf{k}(f_\mathbf{k}^{11}+f_\mathbf{k}^{22}+1)
\,,\label{diff1}
\\
i\partial_t f_\mathbf{k}^{11}=i\partial_t f_\mathbf{k}^{22}
&=&
\sqrt{2}JT_\mathbf{k}(f^{12}_\mathbf{k}-f^{21}_\mathbf{k})
\label{diff2}
\,.
\end{eqnarray}
The last equation implies an effective particle-hole symmetry 
$f_\mathbf{k}^{11}=f_\mathbf{k}^{22}$ valid only in the first order of $1/Z$.
With this symmetry, any stationary state (including the ground state)
with $\partial_tf_\mathbf{k}^{ab}=0$ must obey the condition 
\begin{eqnarray}
\label{stat}
f_\mathbf{k}^{12}=f_\mathbf{k}^{21}
=
\frac{\sqrt{2}JT_\mathbf{k}(2f_\mathbf{k}^{11}+1)}{U-3 JT_\mathbf{k}}
\,.
\end{eqnarray}
Equations (\ref{diff0})-(\ref{diff2}) allow several stationary solutions.
In order to find the ground state one supplementary condition has to be 
imposed. 
One way is to envisage an adiabatic switching procedure starting from the 
exactly known ground state at $J=0$ and slowly increasing $J$ until its 
desired final value $J$ is reached. 
The evolution process has to be very slow in order to avoid the population 
of excited states.
The remaining unknown quantity $f_\mathbf{k}^{11}$ is then obtained by 
noticing that, for any time-dependent $J(t)$, the evolution 
equations~(\ref{diff0})-(\ref{diff2})
leave the following bilinear quantity invariant:
\begin{eqnarray}
\label{inv}
\partial_t
\left[
f_\mathbf{k}^{11}(f_\mathbf{k}^{11}+1)-f_\mathbf{k}^{12}f_\mathbf{k}^{21}
\right]
=0
\,.
\end{eqnarray}
Thus, starting in the Mott state~(\ref{Mott-state}) at zero hopping rate 
$J=0$ with vanishing correlations $f_\mathbf{k}^{ab}(t=0)=0$, we get the 
additional condition
\begin{eqnarray}
\label{inv0}
f_\mathbf{k}^{11}(f_\mathbf{k}^{11}+1)=f_\mathbf{k}^{12}f_\mathbf{k}^{21}
\end{eqnarray}
for all times $t>0$.
Ergo, Eqs.~(\ref{stat}) and (\ref{inv0}) yield 
\begin{eqnarray}
\label{f-solution}
f_\mathbf{k}^{11}
=
\frac{U-3JT_\mathbf{k}-\omega_\mathbf{k}}{2\omega_\mathbf{k}}
\;,\quad
f_\mathbf{k}^{12}
=
\frac{\sqrt{2}JT_\mathbf{k}}{\omega_\mathbf{k}}
\;,
\end{eqnarray}
where
\begin{eqnarray}
\label{eigen-frequency}
\omega_\mathbf{k}
&=&
\sqrt{U^2-6 J UT_\mathbf{k}+J^2 T_\mathbf{k}^2}
\,.
\end{eqnarray}
corresponds to the non-trivial eigenfrequency of the homogeneous part of 
Eqs.~(\ref{diff0})-(\ref{diff2}). 
This expression~(\ref{eigen-frequency}) 
has already been derived using different methods, such as 
the time dependent Gutzwiller approach~\cite{KN2011}, 
the random phase approximation~\cite{SD2005}, or the 
slave boson approach~\cite{HABB2007}, where 
$\omega_\mathbf{k}=\omega_{\mathbf{k}}^{\rm d}-\omega_{\mathbf{k}}^{\rm h}$
is given by the difference between the doublon and holon frequencies. 
Note that this expression~(\ref{eigen-frequency}) differs from the one 
obtained in Ref.~\cite{BPCK12} for a one-dimensional lattice via a 
fermionization approach. 

Thus, the ground-state correlations read (for $\mu\neq\nu$)
\begin{eqnarray}
\label{ground1}
\langle\hat{h}^\dagger_\mu\hat{h}_\nu^{} \rangle_{\rm ground} 
&=&
\langle\hat{p}^\dagger_\mu\hat{p}_\nu^{} \rangle_{\rm ground}
\\
&=&
\frac{1}{N}\sum_\mathbf{k}
\frac{U-3JT_\mathbf{k}-\omega_\mathbf{k}}{2\omega_\mathbf{k}}
e^{i\mathbf{k}\cdot(\mathbf{x}_\mu-\mathbf{x}_\nu)}
\;,
\nonumber
\\
\label{ground2}
\langle\hat{h}^\dagger_\mu\hat{p}_\nu^{} \rangle_{\rm ground} 
&=&
\langle\hat{p}^\dagger_\mu\hat{h}_\nu^{} \rangle_{\rm ground}
\\
&=&
\frac{1}{N}\sum_\mathbf{k}\frac{\sqrt{2}JT_\mathbf{k}}{\omega_\mathbf{k}}
e^{i\mathbf{k}\cdot(\mathbf{x}_\mu-\mathbf{x}_\nu)}
\,.
\nonumber
\end{eqnarray}
Consistent with the (discrete) translational invariance of the lattice, 
these and other two-point correlation functions depend on the distance 
${\bf x}_\mu-{\bf x}_\nu$.
Again, similar results, e.g., the correlator 
$\langle\hat{b}^\dagger_\mu\hat{b}_\nu^{} \rangle_{\rm ground} $
can also be obtained employing other methods, 
such as the the random phase approximation~\cite{SD2005}.
However, the justification of this approximation is another matter -- 
especially for time-dependent situations we are interested in, 
such as a rapidly changing $J(t)$ and the subsequent dephasing of 
quasi-particles etc. 

The above Eqs.~(\ref{ground1}) and (\ref{ground2}) describe the 
correlations and are valid for $\mu\neq\nu$ only. 
The correct on-site density matrix $\rho_\mu$ can be obtained 
from~(\ref{one-site})
which shows that non-vanishing correlations lead 
to small deviations from the lowest-order result $\rho_\mu^0$.
As one would expect, the quantum ground-state fluctuations manifest 
themselves in a small depletion of the unit-filling state 
$\hat\rho_{\mu}^0=\ket{1}_\mu\!\bra{1}$ given by a small but finite 
probability for a particle 
$f_2=\tr\{\hat{\rho}_\mu|2\rangle_\mu\langle 2|\}=
\langle\hat{p}^\dagger_\mu\hat{p}_\mu^{}\rangle$ 
or a hole 
$f_0=\tr\{\hat{\rho}_\mu|0\rangle_\mu\langle 0|\}=
\langle\hat{h}_\mu^{}\hat{h}^\dagger_\mu\rangle$.
To first order in $1/Z$, we get from~(\ref{one-site}) 
\begin{eqnarray}
\label{diff3}
i\partial_t f_0
&=&
i\partial_t f_2
=
\sum_\mathbf{k}\frac{\sqrt{2}J T_\mathbf{k}}{N}
(f_\mathbf{k}^{12}-f_\mathbf{k}^{21})
\nn
&=&
\frac{i}{N}\sum_\mathbf{k} \partial_tf_\mathbf{k}^{11}
\,,
\end{eqnarray}
where we used Eq.~(\ref{diff2}) in the last step.
This equation can be integrated easily and with the initial conditions 
$f_0(t=0)=f_2(t=0)=0$ we find the $1/Z$-corrections to the on-site 
density matrix
\begin{eqnarray}
\label{depletion}
\langle\hat{p}^\dagger_\mu\hat{p}_\mu^{}\rangle 
=
\langle\hat{h}_\mu^{}\hat{h}^\dagger_\mu\rangle 
=
\frac{1}{N}\sum_\mathbf{k}
\frac{U-3JT_\mathbf{k}-\omega_\mathbf{k}}{2\omega_\mathbf{k}}
\,.
\end{eqnarray}
Note that, even though the right-hand side of the above equation looks 
like that of~(\ref{ground1}) for $\mu=\nu$, one should be careful as they 
are derived from two different equations:~(\ref{one-site}) 
and~(\ref{two-sites}).

In an analogous way, we may derive the expression for the ground-state 
energy $E_0$
to first order of $1/Z$, which can be obtained combining Eqs.~(\ref{ground1}), 
(\ref{ground2}) and (\ref{depletion}), and gives
\begin{eqnarray}
\label{onsiteenergy}
\frac{E_0}{N}
=
\sum_\mathbf{k}
\frac{\omega_\mathbf{k}-U}{2N}
+
{\cal O}
\left(
    \frac{1}{Z^2}
\right)
\,.
\end{eqnarray}
This result can also be obtained via the slave boson approach~\cite{HABB2007}
supplemented with the restriction of the Hilbert space to local occupation 
numbers below three.
In our method, this restriction does not have to be put in by hand, but 
follows effectively from the $1/Z$-expansion. 

\section{Quench dynamics}
\label{Quench dynamics}

\begin{figure}[h]
\begin{center}
\includegraphics[width=.9\columnwidth]{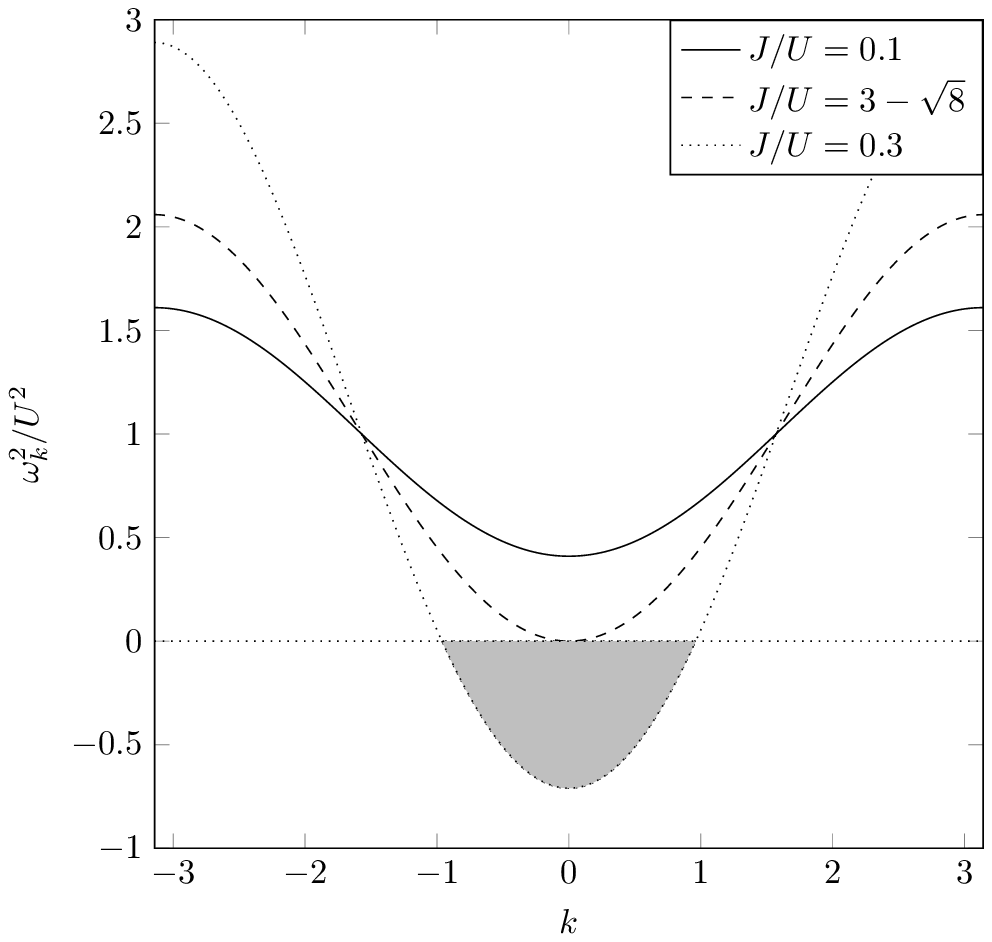}
\includegraphics[width=.9\columnwidth]{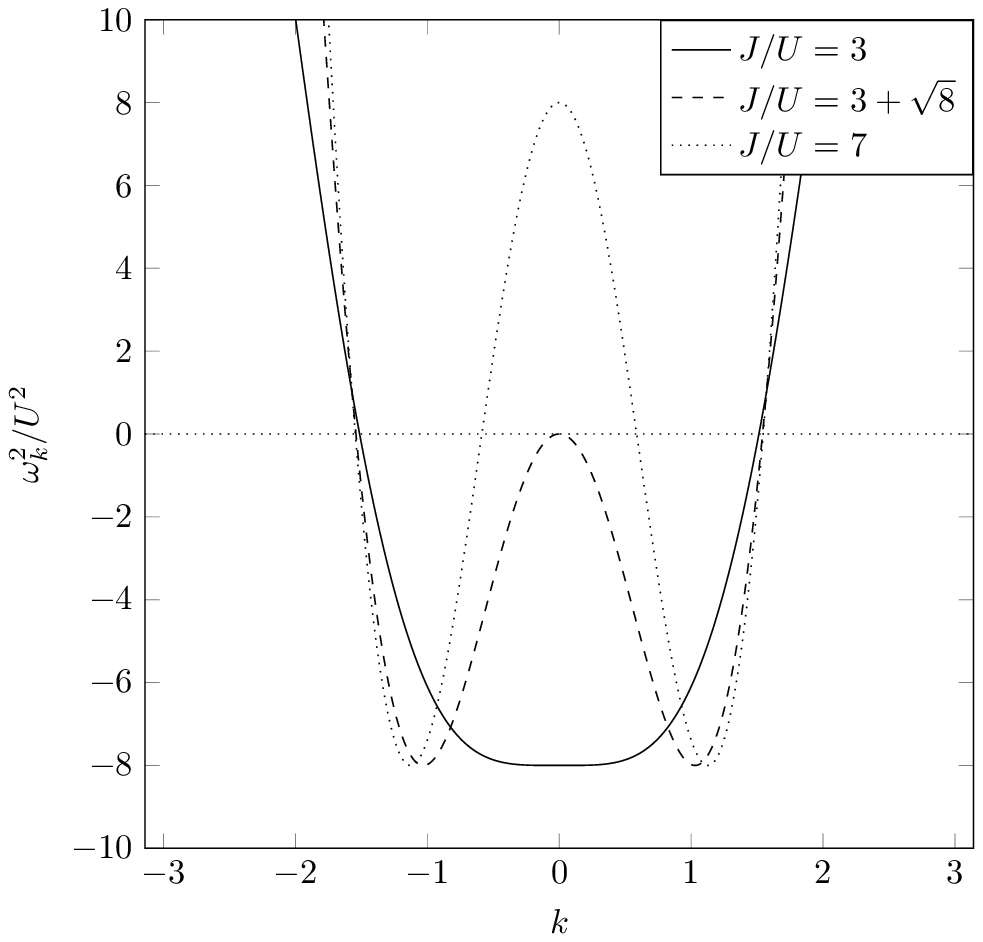}
\end{center}
\caption{Dispersion relation $\omega_k^2/U^2$ in one dimension for 
different values of $J/U$.}
\label{fig-omega}  
\end{figure}

After having studied the ground-state properties of the Mott phase, 
let us consider a quantum quench.
This requires a time-dependent solution of the evolution equations 
(\ref{diff0})-(\ref{diff2}) which crucially depends on the eigenfrequency 
(\ref{eigen-frequency}).
Thus, let us first discuss the general behavior of (\ref{eigen-frequency}).
In view of the definition (\ref{T_k}), $T_{\mathbf{k}}$ adopts its 
maximum value $T_{\mathbf{k}=0}=1$ at $\mathbf{k}=0$.
Thus $\omega_{\mathbf{k}=0}=\Delta\mathcal E$ corresponds to the 
energy gap of the Mott state mentioned in 
Section~\ref{mottsection}. 
For $J=0$, we have a flat dispersion relation $\omega_\mathbf{k}=U$.
If we increase $J$, the dispersion relation $\omega_\mathbf{k}$ 
bends down and the minimum at $\mathbf{k}=0$ approaches the axis. 
Finally, at a critical value of the hopping rate given by 
$J_{\rm c}/U=3-\sqrt{8}\approx 0.17$~\cite{sachdev}
the minimum $\omega_{\mathbf{k}=0}$ touches the axis and thus 
the energy gap vanishes $\Delta\mathcal E=0$.
This marks the transition to the superfluid regime and we can
neither analytically nor adiabatically continue beyond this point. 
However, nothing stops us from suddenly switching $J$ to a final 
value $J_{\rm out}>J_{\rm c}$ beyond this point. 
Of course, this would not be adiabatic anymore and we would 
no longer be close to the ground state.
For hopping rates $J$ which are a bit larger than the critical value 
$J>J_{\rm c}$,
the eigenfrequencies $\omega_\mathbf{k}$ become imaginary for small 
$\mathbf{k}$
indicating an exponential growth of these modes, i.e., an instability. 
This is because the Mott state is no longer the ground state. 
If we consider even larger $J$, we find that the original minimum 
of the dispersion relation $\omega_\mathbf{k}^2$ at $\mathbf{k}=0$
splits into degenerate minima at finite values of $\mathbf{k}$ 
when $J=3U$, while $\mathbf{k}=0$ becomes a local maximum. 
This local maximum even emerges $\omega_{\mathbf{k}=0}^2>0$ 
on the positive side again for $J>U(3+\sqrt{8})$.
Nevertheless, there are always unstable modes for some values of 
$\mathbf{k}$, see Fig.~\ref{fig-omega} and compare~\cite{S11}.

After these preliminaries, let us study a quantum quench from $J=0$
to a finite value $J_{\rm out}<J_{\rm c}$ which is still in the Mott regime. 
For simplicity, we consider a sudden change of $J(t)=J_{\rm out}\Theta(t)$, 
but the calculation can easily be generalized to other scenarios. 
Solving the evolution equations~(\ref{diff0})-(\ref{diff2}) for this case, 
we find~\cite{NS10}
\begin{eqnarray}
\label{quench-h+h}
\langle\hat{h}^\dagger_\mu\hat{h}_\nu^{} \rangle_{\rm quench} 
&=&
\langle\hat{p}^\dagger_\mu\hat{p}_\nu^{} \rangle_{\rm quench}
\\
&=&
\frac{1}{N}\sum_\mathbf{k}
4J^2 T_\mathbf{k}^2\,
\frac{1-\cos(\omega_\mathbf{k}t)}{\omega^2_\mathbf{k}}\,
e^{i\mathbf{k}\cdot(\mathbf{x}_\mu-\mathbf{x}_\nu)}
\;,
\nonumber
\end{eqnarray}
and 
\begin{eqnarray}
\label{quench-h+p}
\langle\hat{h}^\dagger_\mu\hat{p}_\nu^{} \rangle_{\rm quench} 
&=&
\frac{i}{N}\sum_\mathbf{k}
\sqrt{2}JT_\mathbf{k}\,
\frac{\sin(\omega_\mathbf{k}t)}{\omega_\mathbf{k}}\,
e^{i\mathbf{k}\cdot(\mathbf{x}_\mu-\mathbf{x}_\nu)}
\nn
&&
+
\frac{1}{N}\sum_\mathbf{k}
\sqrt{2}JT_\mathbf{k}(U-3JT_\mathbf{k})
\nn
&&
\times
\frac{1-\cos(\omega_\mathbf{k}t)}{\omega_\mathbf{k}^2}\,
e^{i\mathbf{k}\cdot(\mathbf{x}_\mu-\mathbf{x}_\nu)}
\,.
\end{eqnarray}
The remaining correlation can simply be obtained via 
$
\langle\hat{p}^\dagger_\nu\hat{h}_\mu^{}\rangle
=
\langle\hat{h}^\dagger_\mu\hat{p}_\nu^{}\rangle^*
$.
The correlator in terms of the original creation and 
annihilation operators $\hat b_\mu^\dagger$ and $\hat b_\nu$ 
is just a linear combination of these correlation functions 
\bea
\label{quench-b+b}
\langle\hat{b}^\dagger_\mu\hat{b}_\nu^{} \rangle_{\rm quench} 
=
\frac{4JU}{N}\sum_\mathbf{k}T_\mathbf{k}\,
\frac{1-\cos(\omega_\mathbf{k}t)}{\omega^2_\mathbf{k}}\,
e^{i\mathbf{k}\cdot(\mathbf{x}_\mu-\mathbf{x}_\nu)}
\,.
\eea 
The quench $J(t)$ can be realized experimentally by decreasing the 
intensity of the laser field generating the optical lattice 
(which lowers the potential barrier for tunneling and thus 
increases $J$). 
Thus the above prediction should be testable in experiments. 

Note that the same expression would apply to a quench from the 
Mott to the superfluid regime, cf.~\cite{NS10}. 
As explained above, in this case the frequencies $\omega_\mathbf{k}$ 
become imaginary for some $\mathbf{k}$ and thus these modes grow 
exponentially. 
As a result, the expectation value will quickly be dominated by these 
fast growing modes and so most of the details of the initial state 
will become unimportant. 
Of course, this exponential growth cannot continue forever -- 
after some time, the $1/Z$-expansion breaks down since the quantum 
fluctuation are too strong and the growth will saturate. 

\section{Equilibration}
\label{Equilibration}

\begin{center}
\begin{figure}[h]
\includegraphics{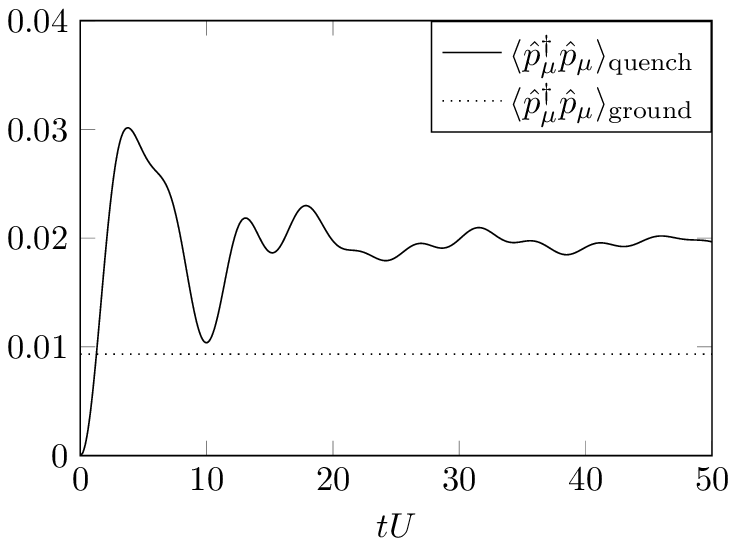}
\includegraphics{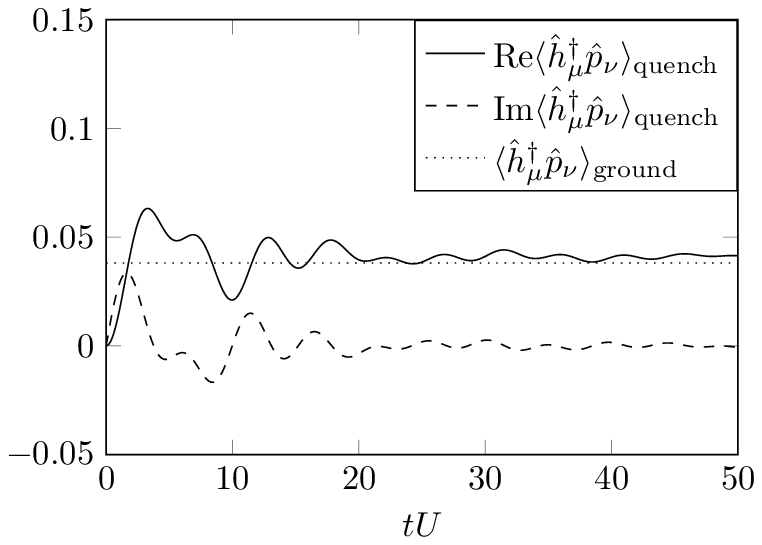}
\begin{center}
\includegraphics{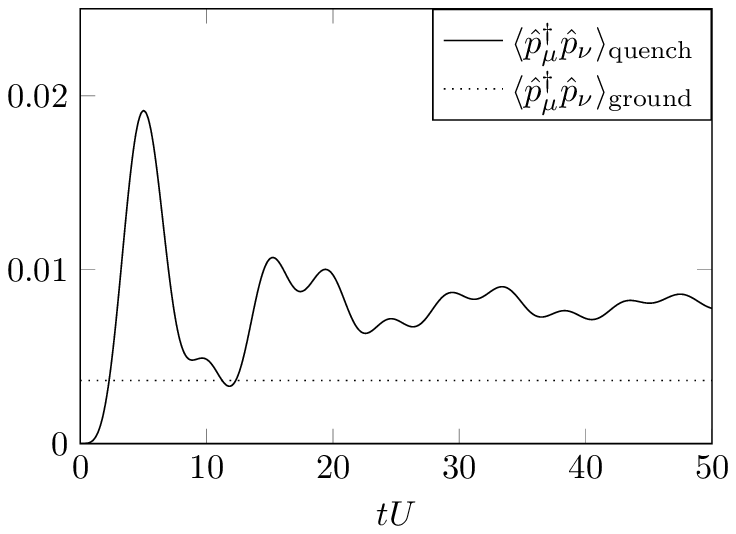}
\end{center}
\caption{Time-dependence of the depletion 
$\langle\hat{p}_{\mu}^\dagger\hat{p}_{\mu}\rangle$ 
and the nearest-neighbor correlations functions 
$\langle\hat{h}_{\mu}^\dagger\hat{p}_{\nu}\rangle$
and 
$\langle\hat{p}_{\mu}^\dagger\hat{p}_{\nu}\rangle$
in three dimensions after the quench within the Mott phase
$J/U=0\to 0.14$ in comparison to their ground-state values.  
After quasi-equilibration, 
$\langle\hat{p}_{\mu}^\dagger\hat{p}_{\nu}\rangle_\mathrm{quench}$ and 
$\langle\hat{p}_{\mu}^\dagger\hat{p}_{\nu}\rangle_\mathrm{ground}$
as well as 
$\langle\hat{p}_{\mu}^\dagger\hat{p}_{\mu}\rangle_\mathrm{quench}$ and 
$\langle\hat{p}_{\mu}^\dagger\hat{p}_{\mu}\rangle_\mathrm{ground}$ 
differ roughly by a factor of two.}
\label{equilbose}
\end{figure}
\end{center}

However, in the following, we shall study a quench within the Mott regime. 
In this case, all frequencies are real $\omega_\mathbf{k}\in\mathbb R$ 
and thus there is no exponential growth -- all modes oscillate.
For an infinite (or at least extremely large) lattice, the oscillations 
in (\ref{quench-h+h}-\ref{quench-b+b}) average out for sufficiently large 
times $t$ and thus these observables approach a quasi-equilibrium value 
\begin{eqnarray}
\label{equil-h+h}
\langle\hat{h}^\dagger_\mu\hat{h}_\nu^{} \rangle_{\rm equil} 
&=&
\langle\hat{p}^\dagger_\mu\hat{p}_\nu^{} \rangle_{\rm equil} 
\\
&=&
\frac{1}{N}\sum_\mathbf{k}
\frac{4J^2 T_\mathbf{k}^2}{\omega^2_\mathbf{k}}\,
e^{i\mathbf{k}\cdot(\mathbf{x}_\mu-\mathbf{x}_\nu)}
\;,
\nonumber
\\
\label{equil-h+p}
\langle\hat{h}^\dagger_\mu\hat{p}_\nu^{} \rangle_{\rm equil} 
&=&
\langle\hat{p}^\dagger_\mu\hat{h}_\nu^{} \rangle_{\rm equil} 
\\
&=&
\frac{1}{N}\sum_\mathbf{k}
\sqrt{2}JT_\mathbf{k}\,
\frac{U-3JT_\mathbf{k}}{\omega_\mathbf{k}^2}
e^{i\mathbf{k}\cdot(\mathbf{x}_\mu-\mathbf{x}_\nu)}
\,.
\nonumber
\end{eqnarray}
The quasi-equilibrium values for the local (on-site) particle and hole 
probabilities can be derived in complete analogy to the previous case
\begin{eqnarray}
\label{parthole}
\langle\hat{p}^\dagger_\mu\hat{p}_\mu^{}\rangle_{\rm equil}  
=
\langle\hat{h}_\mu^{}\hat{h}^\dagger_\mu\rangle_{\rm equil}  
=
\frac{1}{N}\sum_\mathbf{k}
\frac{4J^2 T_\mathbf{k}^2}{\omega^2_\mathbf{k}}
\,.
\end{eqnarray}
Again, it turn out that the result coincides with Eq.~(\ref{equil-h+h}) 
after setting $\nu=\mu$.
For the explicit example of a Bose-Hubbard model on a three-dimensional 
cubic lattice after a quench according to $J/U=0\to 0.14$, the time 
dependences from Eqs.~(\ref{quench-h+h}) and (\ref{quench-h+p}) are plotted 
in Fig.~\ref{equilbose}. 

Having found that the observables considered above approach a 
quasi-equilibrium state, it is natural to ask the question of 
thermalization.
As explained in the Introduction, this is one of the major unsolved 
questions (or rather a set of questions) in quantum many-body 
theory~\cite{KWE11,GME11,EHKKMWW09,KISD11,BCH11}.
Even though we cannot settle this question here, we can compare the 
quasi-equilibrium values obtained above with a thermal state.
To this end, we derive the thermal density matrix $\hat\rho_\beta$ 
corresponding to a given temperature $k_{\rm B}T=1/\beta$. 
Using the grand canonical ensemble, the thermal density operator reads 
\begin{eqnarray}
\hat\rho_\beta
=
\frac{e^{-\beta(\hat{H}-\mu \hat{N})}}
{\tr\{e^{-\beta(\hat{H}-\mu \hat{N})}\}}
\,, 
\end{eqnarray}
where chemical potential $\mu$ will be chosen such that the filling 
is equal to unity.
For small values of $J/U$, we can employ
strong-coupling perturbation theory, i.e., an expansion in powers of $J/U$.
It is useful to introduce the operator~\cite{CFMSE08,FCMSE08}
\begin{eqnarray}
\hat R(\beta)
=
e^{\beta\hat{H}_0}e^{-\beta(\hat{H}_0+\hat{H}_1)}
\,,
\end{eqnarray}
where $\hat{H}_0$ is the diagonal on-site part of the grand canonical 
Hamiltonian $\hat{H}-\mu \hat{N}$ and $\hat{H}_1$ is the hopping term. 
This operator satisfies the differential equation
\begin{eqnarray}
\partial_\beta\hat{R}(\beta)=-\hat{H}_1(\beta)\hat{R}(\beta)
\,, 
\end{eqnarray}
where $\hat{H}_1(\beta)=e^{\beta\hat{H}_0} \hat{H}_1 e^{-\beta\hat{H}_0}$.
In analogy to time-dependent perturbation theory, the operator $\hat{R}$ 
can be calculated perturbatively by integrating this equation with respect 
to $\beta$. 
In first-order perturbation expansion (in $J/U$), we have
(see also Ref.~\cite{CFMSE08}) 
\begin{eqnarray}
\label{thermal-perturbation}
\hat{\rho}_\beta
=
\frac{e^{-\beta\hat{H}_0}}{{\mathfrak Z}_0}
\left(
1
+
\frac{J}{Z}
\sum_{\mu\nu}
T_{\mu\nu}
\,
\hat{b}^\dagger_\mu
\frac{e^{\beta U(\hat{n}_\mu-\hat{n}_\nu)}-1}
{U(\hat{n}_\mu-\hat{n}_\nu)}\,
\hat{b}^{}_\nu
\right)
\end{eqnarray}
with ${\mathfrak Z}_0={\tr\{e^{-\beta \hat{H}_0}\}}$. 
Obviously, the correction to first order in $J/U$ does not affect the 
one-point density matrix $\hat\rho_\mu$ but the two-point correlations.
Thus, we find that the quasi-equilibrium state of the one-point density 
matrix $\hat\rho_\mu$ can indeed be described by a thermal state
provided that we choose the chemical potential as $\mu=U/2$ which gives 
\bea\label{thermdens}
\hat\rho_\mu(\beta)
&\approx&
\frac{e^{-\beta U/2}}{2}\,\ket{0}_\mu\!\bra{0}
+
\left(1-e^{-\beta U/2}\right)\ket{1}_\mu\!\bra{1}
\nn
&&
+
\frac{e^{-\beta U/2}}{2}\,\ket{2}_\mu\!\bra{2}
\,.
\ea
The particular value $\mu=U/2$ of the chemical potential ensures 
that (in first order thermal perturbation theory) we have on average one 
particle per lattice site and the particle-hole symmetry 
$\langle \hat{p}_\mu^\dagger \hat{p}_\mu\rangle=
\langle \hat{h}_\mu^\dagger \hat{h}_\mu\rangle$.
To obtain the correct probabilities, we have to select the temperature 
according to  
\bea
\label{temperature}
e^{-\beta U/2}=2\langle\hat{p}^\dagger_\mu\hat{p}_\mu^{}\rangle_{\rm equil}  
=
\frac{8J^2}{N}\sum_\mathbf{k}
\frac{T_\mathbf{k}^2}{\omega^2_\mathbf{k}}
=\ord(1/Z)
\,,
\ea
which can be deduced from Eqs.~(\ref{parthole}) and (\ref{thermdens}).
Since the depletion is small 
$\langle\hat{p}^\dagger_\mu\hat{p}_\mu^{}\rangle=\ord(1/Z)$,  
we obtain a low effective temperature which scales as $T=\ord(U/\ln Z)$. 
Accordingly, consistent with our $1/Z$-expansion, we can neglect 
higher Boltzmann factors such as $e^{-\beta U}$. 

\section{Correlations}
\label{Correlations}

Of course, the fact that the one-point density matrix $\hat\rho_\mu$ 
can be described (within our limits of accuracy) by a thermal state 
does not imply that the same is true for the correlations.
To study this point, let us calculate the thermal two-point correlator 
from (\ref{thermal-perturbation}). 
To first order in $J/U$ and $1/Z=\ord(e^{-\beta U}/2)$, we find 
\bea
\label{hp-temp-J^1}
\langle\hat{h}^\dagger_\mu\hat{p}_\nu^{} \rangle_\beta
&=&
\langle\hat{p}^\dagger_\mu\hat{h}_\nu^{} \rangle_\beta
=
\frac{\sqrt{2}JT_{\mu\nu}}{ZU}
\nn
&&
+\ord(J^2)+\ord(1/Z^2)
\,,
\ea
while $\langle\hat{h}^\dagger_\mu\hat{h}_\nu^{} \rangle_\beta$
and $\langle\hat{p}^\dagger_\mu\hat{p}_\nu^{} \rangle_\beta$ 
vanish (to first order in $J/U$).
If we compare this to the quasi-equilibrium value 
$\langle\hat{h}^\dagger_\mu\hat{p}_\nu^{} \rangle_{\rm equil}$ in 
(\ref{equil-h+p}), we find that they coincide to first order in $J/U$ 
\bea
\label{hp-equil-J^1}
\langle\hat{h}^\dagger_\mu\hat{p}_\nu^{} \rangle_{\rm equil}
&=&
\langle\hat{p}^\dagger_\mu\hat{h}_\nu^{} \rangle_{\rm equil}
=
\frac{\sqrt{2}JT_{\mu\nu}}{ZU}
\nn
&&
+\ord(J^2)+\ord(1/Z^2)
\,.
\ea
This is perhaps not too surprising since the same value can be obtained 
from the ground-state fluctuations 
$\langle\hat{h}^\dagger_\mu\hat{p}_\nu^{} \rangle_{\rm ground}=
\langle\hat{p}^\dagger_\mu\hat{h}_\nu^{} \rangle_{\rm ground}$  
in (\ref{ground2}) after expanding them to first order in $J/U$. 
Due to the low effective temperature $T=\ord(U/\ln Z)$, the lowest 
Boltzmann factor is suppressed by $e^{-\beta U/2}=\ord(1/Z)$.
As a consequence, because the correlations are small $\ord(1/Z)$,
their finite-temperature corrections are even smaller $\ord(1/Z^2)$,
and thus can be neglected.  

The same is true for the other correlations 
$\langle\hat{h}^\dagger_\mu\hat{h}_\nu^{} \rangle=
\langle\hat{p}^\dagger_\mu\hat{p}_\nu^{} \rangle$.
All of them: the ground-state correlators
$\langle\hat{h}^\dagger_\mu\hat{h}_\nu^{} \rangle_{\rm ground}=
\langle\hat{p}^\dagger_\mu\hat{p}_\nu^{} \rangle_{\rm ground}$ 
in (\ref{ground1}), the quasi-equilibrium correlators 
$\langle\hat{h}^\dagger_\mu\hat{h}_\nu^{} \rangle_{\rm equil}=
\langle\hat{p}^\dagger_\mu\hat{p}_\nu^{} \rangle_{\rm equil}$ 
in (\ref{equil-h+h}), as well as the thermal correlators 
$\langle\hat{h}^\dagger_\mu\hat{h}_\nu^{} \rangle_\beta$
and $\langle\hat{p}^\dagger_\mu\hat{p}_\nu^{} \rangle_\beta$ 
vanish to first order in $J/U$.
Therefore, to first order in $J/U$ and $1/Z$, the thermal state 
can describe the observables under consideration.
However, going to the next order in $J$, this description breaks down.
This failure can even be shown without explicitly calculating 
$\hat R(\beta)$ up to second order. 
If we compare the quasi-equilibrium correlators~(\ref{equil-h+h}) 
\bea
\label{hp-equil-J^2}
\langle\hat{h}^\dagger_\mu\hat{h}_\nu^{} \rangle_{\rm equil}
&=&
\langle\hat{p}^\dagger_\mu\hat{p}_\nu^{} \rangle_{\rm equil}
=
\frac{4J^2 }{U^2Z^2}\sum_\kappa T_{\mu\kappa}T_{\kappa\nu}
\nn
&&
+\ord(J^3)+\ord(1/Z^2)
\,,
\ea
with the ground-state correlations in (\ref{ground1}),
expanded to the same order in $J$  
\bea
\label{hp-temp-J^2}
\langle\hat{h}^\dagger_\mu\hat{h}_\nu^{} \rangle_{\rm ground}
&=&
\langle\hat{p}^\dagger_\mu\hat{p}_\nu^{} \rangle_{\rm ground}
=
\frac{2J^2 }{U^2Z^2}\sum_\kappa T_{\mu\kappa}T_{\kappa\nu}
\nn
&&
+\ord(J^3)+\ord(1/Z^2)
\,,
\ea
we find a discrepancy by a factor of two~\cite{footnote-lattice-structure}. 
I.e., after the quench, these correlations settle down to a value 
which is twice as large as in the ground state (see Fig.~\ref{equilbose}).
This factor of two has already been found elsewhere in the context 
of standard time-dependent and time-independent perturbation theory, 
see also~\cite{MK09}.
This is incompatible with the small Boltzmann factors 
$e^{-\beta U/2}=\ord(1/Z)$ and would require a comparably large 
effective temperature $T=\ord(U)$ instead of $T=\ord(U/\ln Z)$.
However, such a large effective temperature $T=\ord(U)$ is 
inconsistent with the small on-site depletion~(\ref{temperature}).

This distinction between local observables 
(which become approximately thermal) and non-local correlations 
(which are incompatible with this thermal state) has already been observed 
in other scenarios using different approaches.
For the Bose-Hubbard model, quenches from the superfluid phase to the 
Mott state at finite values of $J$ and $U$ have been studied in Ref.~\cite{KLA07}, 
where a significant dependence on the final values of $J$ and $U$ has been 
observed: 
For large values of the final $U$, the (quasi) equilibrated correlations 
deviate significantly from a thermal state, whereas this deviation is not 
pronounced for smaller values. 
In contrast, a quench between the two exactly solvable cases $J=0$ on the 
one hand and $U=0$ on the other hand has been studied in Ref.~\cite{CDEO08}.
This case can be solved exactly and consistent with the existence of the 
conservation laws (as mentioned in the Introduction), only partial 
thermalization is observed. 
Further studies have been devoted to bosonic superlattices 
(see, e.g., Ref.~\cite{CFMSE08}) and fermionic systems
(see the discussion at the end of Section~\ref{fermi-noneq}), for example. 
Unfortunately, a general and unifying understanding of all these 
non-equilibrium phenomena is still missing.  

\section{Second Order in $1/Z$}
\label{Z2}

So far, we have only considered the first order in $1/Z$.
Now let us discuss the effect of higher orders by means of a few examples.
Unfortunately, the complete derivation is rather lengthy and cannot be given 
here, it will be presented elsewhere~\cite{elsewhere}. 

Let us go back to the derivation from (\ref{two-sites}) to 
(\ref{two-sites-approx}) and include $1/Z^2$ corrections. 
To achieve this level of accuracy, we should not replace the exact 
one-point density matrix $\hat\rho_\mu$ by it lowest-order approximation 
$\hat\rho_\mu^0$ but include its first-order corrections in 
(\ref{depletion}), i.e., the quantum depletion 
$f_0=\tr\{\hat{\rho}_\mu|0\rangle_\mu\langle 0|\}=
\langle\hat{h}_\mu^{}\hat{h}^\dagger_\mu\rangle=\ord(1/Z)$ 
of the unit filling (Mott) state in Eq.~(\ref{depletion}). 
This results in a renormalization of the eigenfrequency 
\begin{eqnarray}
\label{renomega}
\omega_\mathbf{k}^{\rm ren}
=
\sqrt{U^2-6JT_\mathbf{k}(1-3f_0)+J^2T_\mathbf{k}^2(1-3f_0)^2}
\,,
\end{eqnarray}
Since the net effect can roughly be understood as a reduction of the 
effective hopping rate $J^{\rm ren}=J(1-3f_0)$, it is easy to visualize 
that this implies also a decrease of the effective propagation velocity.

There are also other $1/Z^2$ corrections in (\ref{two-sites-approx}) such as
the three-point correlator $\hat\rho^{\rm corr}_{\mu\nu\kappa}$ but they act
as source terms and do not affect the eigenfrequency (at second order).
However, there are other quantities where these source terms are crucial.
In particular, we consider two-point correlation functions which vanish to 
first order in $1/Z$, in contrast to contributions such as 
$\langle\hat{b}^\dagger_\mu\hat{b}_\nu^{} \rangle$ discussed above.
One important example is the particle-number correlation, i.e., 
$\langle\hat{n}_\mu\hat{n}_\nu\rangle-
\langle\hat{n}_\mu\rangle\langle\hat{n}_\nu\rangle$. 
After a somewhat lengthy calculation, we find for the ground-state 
correlations
\begin{eqnarray}\label{number}
\langle\hat{n}_\mu\hat{n}_\nu\rangle-
\langle\hat{n}_\mu\rangle\langle\hat{n}_\nu\rangle=
\nn
\frac{2}{N^2}\sum_{\mathbf{p},\mathbf{q}}
e^{i(\mathbf{p}+\mathbf{q})\cdot(\mathbf{x}_\mu-\mathbf{x}_\nu)}
\left(f^{11}_\mathbf{p}f^{11}_\mathbf{q}-f^{12}_\mathbf{p}f^{21}_\mathbf{q}
\right)
\,,
\end{eqnarray}
where $f^{12}_\mathbf{p},f^{21}_\mathbf{p}$ and 
$f^{11}_\mathbf{p}$ are given through the relations (\ref{f-solution}).
Note that the above result is non-perturbative in $J/U$, see, for example, 
the non-polynomial dependence of $\omega_\mathbf{k}$ on $J$. 

As a related example, the parity correlator reads
\begin{eqnarray}
\label{parity}
\langle(-1)^{\hat{n}_\mu}(-1)^{\hat{n}_\nu}\rangle-
\langle(-1)^{\hat{n}_\mu}\rangle\langle(-1)^{\hat{n}_\mu}\rangle
=
\nn
\frac{8}{N^2}\sum_{\mathbf{p},\mathbf{q}}
e^{i(\mathbf{p}+\mathbf{q})\cdot(\mathbf{x}_\mu-\mathbf{x}_\nu)}
\left(f^{11}_\mathbf{p}f^{11}_\mathbf{q}+f^{12}_\mathbf{p}f^{21}_\mathbf{q}
\right)
\,.
\end{eqnarray}
In analogy to the previous Section, we can also study the correlations 
after a quantum quench with $J(t)=J\Theta(t)$.
Again, there are no contributions to the particle-number and parity 
correlations in first order $1/Z$ -- but, to second order $1/Z$, 
we find formally the same expressions as in the static case (\ref{number}) 
and (\ref{parity}) where $f^{12}_\mathbf{p}(t)$, $f^{21}_\mathbf{p}(t)$, 
and $f^{11}_\mathbf{p}(t)$ are now given by equations (\ref{quench-h+h}) 
and (\ref{quench-h+p}).
The parity correlations after a quench have been experimentally observed
in a one-dimensional setup~\cite{CBPE12}.
Although the hierarchical expansion relies on a large coordination number,
we find qualitative agreement between the theoretical prediction 
(\ref{parity}) for $Z=2$ and the results from~\cite{CBPE12}.
For large times $t$ and distances $\mathbf{x}_\mu-\mathbf{x}_\nu$, 
we may estimate the integrals over $\mathbf{p}$ and $\mathbf{q}$ in 
the expressions (\ref{number}) and (\ref{parity}) via the 
stationary-phase or saddle-point approximation.
The dominant contributions stem from the momenta satisfying the 
saddle-point condition
\begin{eqnarray}
\label{statphase}
\nabla_\mathbf{k}
\left[
\mathbf{k}\cdot(\mathbf{x}_\mu-\mathbf{x}_\nu)\pm\omega_\mathbf{k}t
\right]
=0
\,.
\end{eqnarray}
Thus their structure is determined by the group velocity 
$\mathbf{v}_\mathbf{k}=\nabla_\mathbf{k}\omega_\mathbf{k}$. 
If the equation $\mathbf{x}_\mu-\mathbf{x}_\nu=\pm\mathbf{v}_\mathbf{k}t$ 
has a real solution $\mathbf{k}$, i.e., if the distance 
$\mathbf{x}_\mu-\mathbf{x}_\nu$ can be covered in the time $t$ with the 
group velocity $\mathbf{v}_\mathbf{k}$, then we get a stationary-phase
solution -- otherwise the integral will be exponentially suppressed 
(i.e., the saddle point $\mathbf{k}$ becomes complex). 
For a given direction in ${\bf k}$-space, the maximum group velocity determines the maximum propagation speed of correlations,
i.e., the effective light cone. In a hypercubic lattice in $D$ dimensions with small $J$, for example,
it is given by $v_{\rm max}\approx 3J/D$ along the lattice axes and by $v_{\rm max}\approx 3J/\sqrt{D}$
along the diagonal (where all the components of ${\bf v}_{\rm max}$ are equal to each other).
A similar result has been obtained in Ref.~\cite{BPCK12} for the one-dimensional 
Bose-Hubbard model. 
For an experimental realization, see, e.g., Ref.~\cite{CBPE12}.

\section{Exact numerical results}
\label{Numerical}

\begin{figure}[ht]

%
%
%

\includegraphics[width=7.5cm]{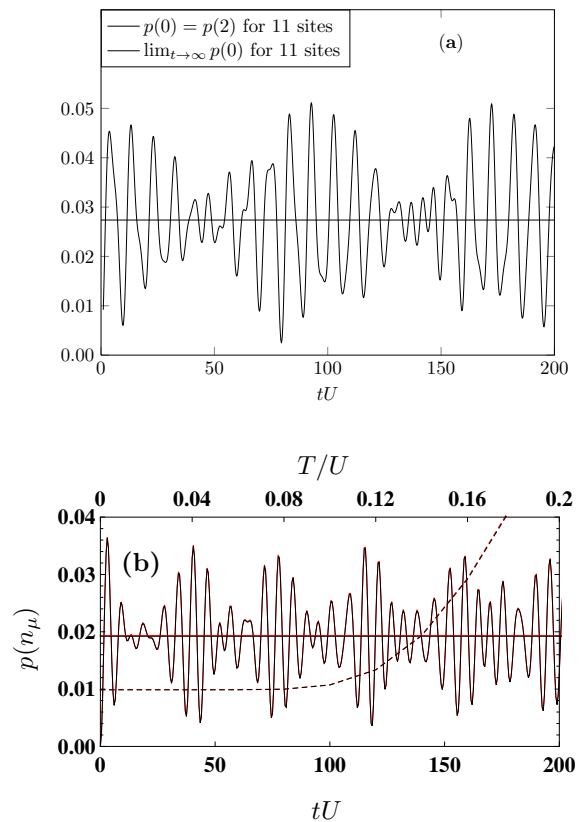}

\caption
{
(Color online)
Time evolution of the probabilities to have zero 
(red) and two (black) particles after quench from $J/U=0$ to $J/U=0.1$
in a one-dimension lattice of $11$ sites with $n=1$ atom per site.
Straight horizontal lines show the values averaged over an infinite 
evolution time.
Top {\bf (a)}: first order of $1/Z$ expansion, see Eqs.~(\ref{quench-h+h}),
~(\ref{equil-h+h}).
Bottom {\bf (b)}: exact diagonalization.
Dashed lines: Probabilities to have $n_\mu=0$ (red), $2$ (black) atoms in a 
thermal state at $J/U=0.1$ as a function of temperature $T$.
Note the different scales for the time $t$ and temperature $T$ dependences.
}
\label{pn-1D-t}
\end{figure}

\begin{figure}[ht]

%
%
%

\includegraphics[width=7.7cm]{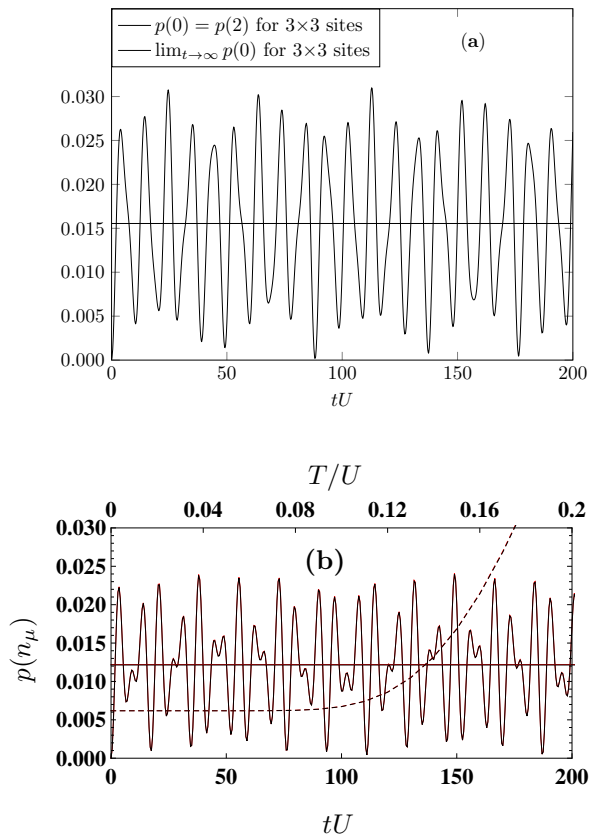}

\caption
{
(Color online) Time evolution of the probabilities to have zero 
(red) or two (black) particles after quench from $J/U=0$ to $J/U=0.1$
in a two-dimension lattice of $3\times 3$ sites with $n=1$ atom per site.
Straight horizontal lines show the values averaged over an infinite 
evolution time.
Top {\bf (a)}: first order of $1/Z$ expansion, see Eqs.~(\ref{quench-h+h}),
~(\ref{equil-h+h}).
Bottom {\bf (b)}: exact diagonalization.
Dashed lines: Probabilities to have $n_\mu=0$ (red), $2$ (black) atoms in a 
thermal state at $J/U=0.1$ as a function of temperature $T$.
Note the different scales for the time $t$ and temperature $T$ dependences.
}
\label{pn_2D-t}
\end{figure}

In order to test the quality of our $1/Z$ expansion, we compare the 
predictions of our first-order calculations with exact numerical results 
for the probabilities
$p(n_\mu)=\bra{n}\hat\rho_\mu\ket{n}$ and correlation functions 
$\langle\hat{b}^\dagger_\mu\hat{b}_\nu^{} \rangle$ in one- and two-dimensional 
finite lattices. 
They are obtained by full diagonalization of the Bose-Hubbard Hamiltonian 
with periodic boundary conditions without any truncation of the Hilbert space.
This allows us to calculate exactly the complete time evolution of any 
quantity as well as their mean values averaged over an infinite time.
The initial state can be arbitrary and in the calculations presented below
it was chosen in the form described by Eq.~(\ref{init-state}).
The full diagonalization provides also a possibility of exact calculations 
of the thermal averages.

The time evolution of the probabilities $p(n_\mu=0)$ and $p(n_\mu=2)$, 
which are by definition equivalent to the quantities
$\langle\hat{h}_\mu^{}\hat{h}^\dagger_\mu\rangle$ and
$\langle\hat{p}^\dagger_\mu\hat{p}_\mu^{}\rangle$
considered in the previous sections, is shown in
Figs.~\ref{pn-1D-t} and~\ref{pn_2D-t} for one- and two-dimensional 
lattices, respectively.
Due to finite-size effects (see also Ref.~\cite{GR10}),  
these probabilities oscillate around their  
averaged values shown by straight horizontal lines.
For the chosen value of $J/U=0.1$, the behavior of $p(0)$ is almost 
indistinguishable from that of $p(2)$, consistent with the $1/Z$-expansion. 
The probabilities for thermal equilibrium states corresponding to the 
final value of $J/U$ (depending on their temperature $T$) 
are also plotted for comparison. 
We observe that the time-averaged values of the probabilities correspond to 
an effective temperature of about $0.14~U$, which is consistent with the 
results of Sec.~\ref{Equilibration}. 
Furthermore, we find that, in a one-dimensional lattice, our $1/Z$-approach 
underestimates the typical frequency scales and overestimates the 
characteristic amplitudes of the probabilities by roughly the same factor 
of $\approx1.4$. 
This might be an indication of the effective renormalization of the hopping 
rate $J^{\rm ren}=J(1-3f_0)$ by the quantum fluctuations discussed in the 
previous section (which are neglected to first order in $1/Z$). 
In two dimensions, this discrepancy is still present -- albeit noticeably
smaller. 
In total, we see that the quantum fluctuations in two dimensions are 
smaller than in one dimension -- and that our $1/Z$-expansion becomes 
better (as one would expect). 

\begin{figure}

%
%
%

\includegraphics[width=7.85cm]{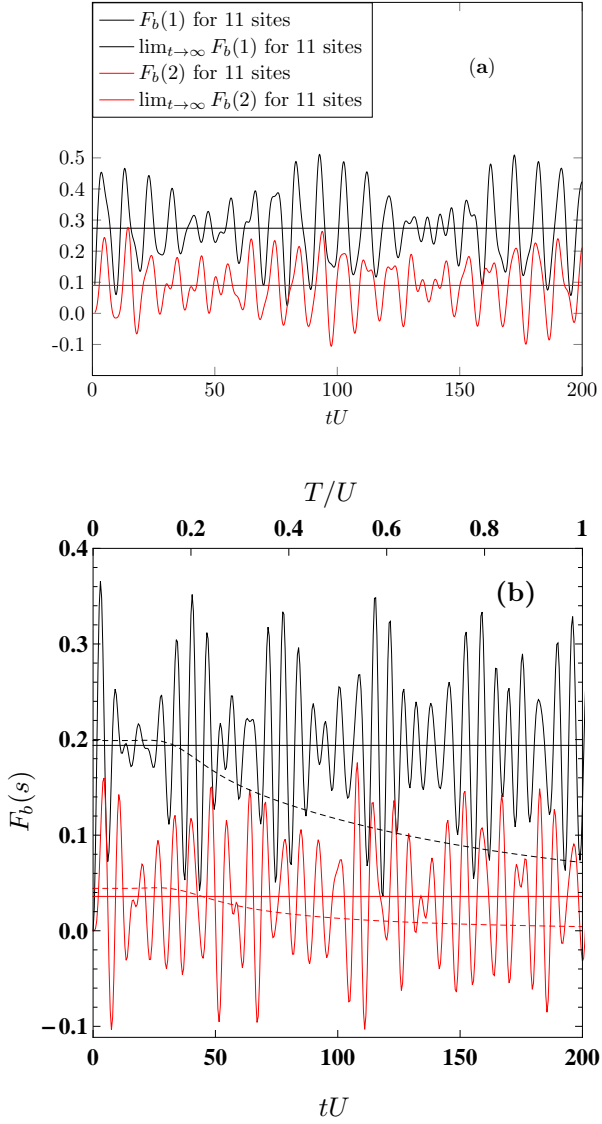}

\caption
{
(Color online)
Correlation function 
$F_b(s)=\langle\hat{b}^\dagger_\mu\hat{b}_\nu^{} \rangle$ for nearest 
neighbors $s=1$~[the solid black (upper) curve] and next-to-nearest neighbors $s=2$~[the solid red (lower) curve]
after a quench from $J/U=0$ to $J/U=0.1$ in a one-dimension lattice of $11$ sites with $n=1$ atom per site.
Straight horizontal lines show the values averaged over an infinite evolution time.
Top {\bf (a)}: first order of $1/Z$ expansion, see Eqs.~(\ref{quench-b+b}),
~(\ref{equil-h+h}),~(\ref{equil-h+p}).
Bottom {\bf (b)} exact diagonalization.
Dashed lines in panel {\bf (b)}:
$F_b(1)$~[black (upper)] and $F_b(2)$~[red (lower)] in a thermal state at $J/U=0.1$ as functions of temperature $T$.
Note the different scales for the time $t$ and temperature $T$ dependences.
}
\label{obdm-1D-t}
\end{figure}

\begin{figure}

%
%
%

\includegraphics[width=8.05cm]{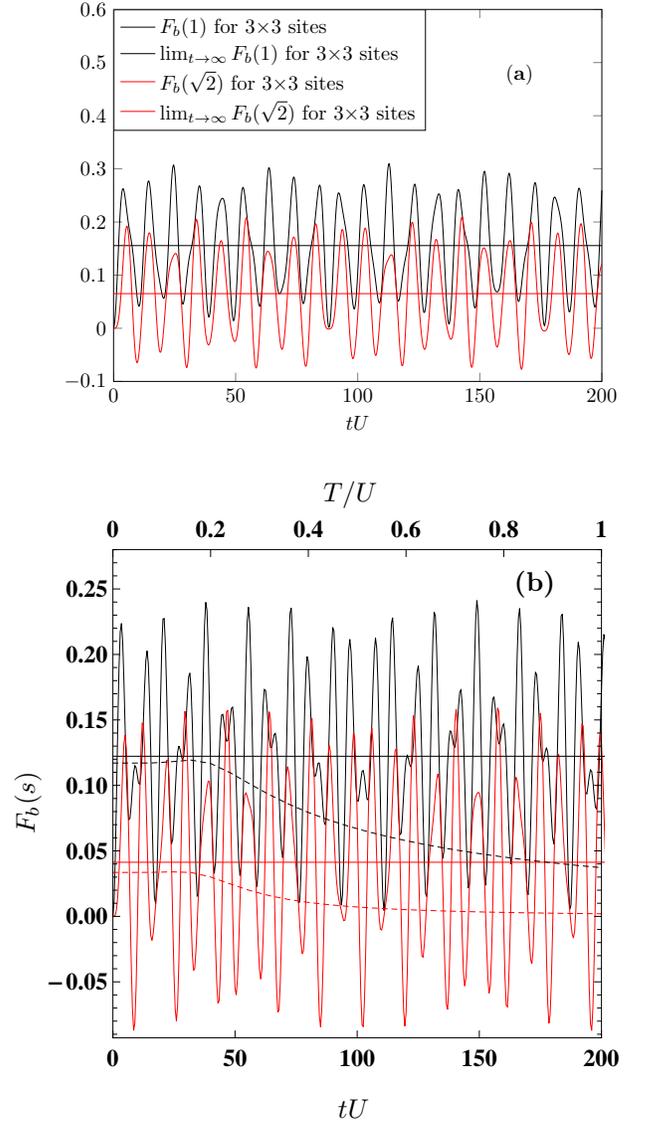}

\caption{
(Color online)
Correlation function 
$F_b(s)=\langle\hat{b}^\dagger_\mu\hat{b}_\nu^{} \rangle$ 
for nearest neighbors $s=1$~[the solid black (upper) curve] and next-to-nearest (diagonal) 
neighbors $s=\sqrt{2}$~[the solid red (lower) curve] after a quench from $J/U=0$ to $J/U=0.1$
in a two-dimension lattice of $3\times 3$ sites with $n=1$ atom per site.
Straight horizontal lines show the values averaged over an infinite evolution time.
Top {\bf (a)}: first order of $1/Z$ expansion, see Eqs.~(\ref{quench-b+b}),
~(\ref{equil-h+h}),~(\ref{equil-h+p}).
Bottom {\bf (b)}: exact diagonalization.
Dashed lines in panel {\bf (b)}: $F_b(1)$~[black (upper)] and $F_b(\sqrt{2})$~[red (lower)]
in a thermal state at $J/U=0.1$ as functions of temperature $T$.
Note the different scales for the time $t$ and temperature $T$ dependences.
}
\label{obdm-2D-t}
\end{figure}

The time dependence of the correlation functions 
$\langle\hat{b}^\dagger_\mu\hat{b}_\nu^{} \rangle$
presented in Figs.~\ref{obdm-1D-t},~\ref{obdm-2D-t} 
displays similar oscillating character and the comparison of the 
$1/Z$-expansion with exact diagonalization reveals the same
characteristic features.
In the one-dimensional lattice, their time-averaged values can again be 
approximately described by an effective temperature of about $0.2~U$, 
but this temperature is already significantly larger than that for the 
probabilities $p(n_\mu)$. 
In contrast, in the two-dimensional case, the time-averaged correlation 
functions cannot be described at all by a thermal state, 
see Fig.~\ref{obdm-2D-t} since are larger than the thermal correlations 
at any temperature. 
This failure of an effective temperature in the two-dimensional system is 
consistent with the result obtained within the $1/Z$-expansion in 
Sec.~\ref{Equilibration}. 
Note that the situation considered here is quite different from a quench 
across the critical point (i.e., Mott-superfluid or superfluid-Mott) 
for which qualitatively different results have been obtained in 
\cite{KLA07,GR10}, for example. 

In general, we come to the conclusion that our $1/Z$-expansion agrees 
qualitatively surprisingly well with exact diagonalization even in one and 
two dimensions, although the values of $1/Z=0.5$ and $0.25$ are not so small.
Furthermore, we observe that the quantitative agreement between our 
$1/Z$-expansion and the numerical results becomes better when going from one 
to two dimensions, as one would expect.

\section{Fermi-Hubbard model}
\label{Fermi-Hubbard Model}

Now, after having studied the bosonic case, let us investigate the 
Fermi-Hubbard model~\cite{H63,EFGKK05,F91}.
We shall find many similarities to the Bose-Hubbard model -- but also crucial 
differences.
The Hamiltonian reads 
\begin{eqnarray}
\label{Fermi-Hubbard-Hamiltonian}
\hat H
=
-\frac{J}{Z}\sum_{\mu\nu,s}T_{\mu\nu}\hat{c}_{\mu,s}^\dagger\hat{c}_{\nu,s}
+U\sum_{\mu}\hat{n}_\mu^\uparrow\hat{n}_\mu^\downarrow
\,.
\end{eqnarray}
The nomenclature is the same as in the bosonic case 
(\ref{Bose-Hubbard-Hamiltonian}) but with an additional spin label $s$ 
which can assume two values $s=\uparrow$ or $s=\downarrow$.
In the following, we consider the case of half-filling 
$\langle\hat{n}_\mu^\uparrow+\hat{n}_\mu^\downarrow\rangle=1$
where half the particles are in the $s=\uparrow$ state and the other 
have $s=\downarrow$.
Note that the total particle numbers  
$\hat N^\uparrow=\sum_\mu \hat n_\mu^\uparrow$ 
and 
$\hat N^\downarrow=\sum_\mu \hat n_\mu^\downarrow$
for each spin species are conserved separately 
$[\hat H,\hat N^\uparrow]=[\hat H,\hat N^\downarrow]=0$. 
The creation and annihilation operators satisfy the fermionic 
anti-commutation relations 
\bea
\label{fermionic-commutation}
\left\{\hat c_{\nu,a},\hat c_{\mu,b}^\dagger\right\}
=\delta_{\mu\nu}\delta_{ab}
\;,\,
\left\{\hat c_{\nu,a},\hat c_{\mu,b}\right\}
=
\left\{\hat c_{\nu,a}^\dagger,\hat c_{\mu,b}^\dagger\right\}
=0
\,.
\ea
The fermionic nature of the particles has important consequences.
For example, let us estimate the expectation value of the hopping 
Hamiltonian $\hat H_J$. 
Introducing the ``coarse-grained'' operator 
\bea
\label{c-sigma}
\hat c_{\mu,s}^\Sigma
=
\frac{1}{\sqrt{Z}}\sum_{\nu}T_{\mu\nu}\hat{c}_{\nu,s}
\,,
\ea
we may write the expectation value of the tunneling energy $\hat H_J$ 
per lattice site for one spin species $s$ as 
$-J\langle\hat c_{\mu,s}^\dagger\hat c_{\mu,s}^\Sigma\rangle/\sqrt{Z}$. 
This expectation value can be interpreted as a scalar product of the 
two vectors $\hat c_{\mu,s}\ket{\Psi}$ and $\hat c_{\mu,s}^\Sigma\ket{\Psi}$
and hence it is bounded by  
\bea
\left|\bra{\Psi}\hat c_{\mu,s}^\dagger\hat c_{\mu,s}^\Sigma\ket{\Psi}\right|
\leq
||\hat c_{\mu,s}\ket{\Psi}||\cdot||\hat c_{\mu,s}^\Sigma\ket{\Psi}||
\,.
\ea
Inserting $||\hat c_{\mu,s}\ket{\Psi}||^2=
\bra{\Psi}\hat c_{\mu,s}^\dagger\hat c_{\mu,s}\ket{\Psi}=
\bra{\Psi}\hat n_{\mu,s}\ket{\Psi}$, we get the expectation value of the 
number operator $\hat n_{\mu,s}$. 
In contrast to the bosonic case, this operator is bounded and thus we find 
$||\hat c_{\mu,s}\ket{\Psi}||\leq1$. 
Furthermore, the operator $\hat c_{\mu,s}^\Sigma$ in (\ref{c-sigma}) obeys 
the same anti-commutation relations (\ref{fermionic-commutation}) and thus 
we find $||\hat c_{\mu,s}^\Sigma\ket{\Psi}||\leq1$ in complete analogy.
Consequently, the absolute value of the tunneling energy per lattice site 
is below $2J/\sqrt{Z}$, i.e., decreases for large $Z$. 

The above result implies that the interaction term $\propto U$ always 
dominates (except in the trivial case $U=0$) in the limit $Z\to\infty$ 
under consideration.
Hence, we are in the strongly interacting Mott regime and do not find 
anything analogous to the Mott--superfluid transition as in the bosonic 
case. 
Note that often~\cite{MV89,F99} a different $Z$-scaling is considered, 
where the hopping term scales with $J/\sqrt{Z}$ instead of $J/Z$ as in 
(\ref{Fermi-Hubbard-Hamiltonian}).
Using this $J/\sqrt{Z}$ scaling, one can study the transition from the 
Mott state to a metallic state which is supposed to occur at a critical 
value of $J$ where -- roughly speaking -- the hopping term starts to 
dominate over the interaction term.
However, this transition is not as well understood as the Mott--superfluid 
transition in the bosonic case. 
With our $J/Z$-scaling in (\ref{Fermi-Hubbard-Hamiltonian}), we study a 
different corner of the phase space where we can address question such 
as tunneling in tilted lattices and equilibration vs thermalization etc. 

\subsection{Symmetries and Degeneracy}

In addition to the usual invariances already known from the bosonic case, 
the Fermi-Hubbard model has some more symmetries.
For example, the particle-hole symmetry 
$\hat c_{\mu,s}^\dagger\leftrightarrow\hat c_{\mu,s}$ and thus 
$\hat n_{\mu,s}=\hat c_{\mu,s}^\dagger\hat c_{\mu,s}
\leftrightarrow
\hat{\bar n}_{\mu,s}=\hat c_{\mu,s}\hat c_{\mu,s}^\dagger
=1-\hat n_{\mu,s}$ 
is no longer an effective approximate symmetry, but becomes exact 
(for the case of half-filling considered here). 

Furthermore, there is an effective $SU(2)$-symmetry corresponding to the
spin degrees of freedom. 
To specify this, let us introduce the effective spin operators
\bea
\label{spin-operators}
\hat S_\mu^z
=
\frac12
\sum\limits_{ab}
\hat c_{\mu,a}^\dagger\,
\sigma^z_{ab}\,
\hat c_{\mu,b}
=
\frac12
\left(\hat{n}_\mu^\uparrow-\hat{n}_\mu^\downarrow\right)
\,,
\ea
and analogously 
$\hat S_\mu^x=\sum_{ab}\hat c_{\mu,a}^\dagger\sigma^x_{ab}\hat c_{\mu,b}/2$
as well as 
$\hat S_\mu^y=\sum_{ab}\hat c_{\mu,a}^\dagger\sigma^y_{ab}\hat c_{\mu,b}/2$
where $\sigma^{x,y,z}_{ab}$ are the usual Pauli spin matrices.
These operators satisfy the usual spin, i.e., $SU(2)$, commutation relations 
and the Fermi-Hubbard Hamiltonian (\ref{Fermi-Hubbard-Hamiltonian})
is invariant under global $SU(2)$ rotations generated by the total spin 
operators $\hat{\mathbf S}_{\rm tot}=\sum_\mu\hat{\mathbf S}_\mu$.

In the case of zero hopping $J=0$, this global $SU(2)$ invariance even
becomes a local symmetry, i.e., we may perform a spin rotation at each 
site without changing the energy.
As a result, the ground state (at half filling) is highly degenerate for
$J=0$ in contrast to the Bose-Hubbard model (at integer filling). 
This degeneracy can be lifted by an additional staggered magnetic field 
(see \ref{groundcorr}) and is related to the spin modes which become 
arbitrarily soft for small $J$. 
In this limit $J\ll U$, their dynamics can be described by an effective 
Hamiltonian, which is basically the Heisenberg model 
\begin{eqnarray}
\label{Heisenberg-Hamiltonian}
\hat H
=
\frac{2J^2}{Z^2U}
\sum_{\mu\nu}T_{\mu\nu}\,\hat{\mathbf S}_\mu\cdot\hat{\mathbf S}_\nu
\,,
\end{eqnarray}
with an effective anti-ferromagnetic coupling constant
of order $1/Z^2$. 
This effective Hamiltonian describes the Fermi-Hubbard Hamiltonian 
(\ref{Fermi-Hubbard-Hamiltonian}) for half-filling in the low-energy 
sub-space where we have one particle per site, but with a variable spin
$\hat{\mathbf S}_\mu$. 

In order to avoid complications such as frustration for the 
anti-ferromagnetic Heisenberg model (\ref{Heisenberg-Hamiltonian}), 
we assume a bipartite lattice -- i.e., we can divide the total lattice
into two sub-lattices $\cal A$ and $\cal B$ such that, for each site in 
$\mu\in\cal A$, all the neighboring sites $\nu$ belong to $\cal B$ 
and {\em vice versa}.
In this case, the ground state of the Heisenberg model 
(\ref{Heisenberg-Hamiltonian}) approaches the N\'eel state for large $Z$
\begin{eqnarray}
\label{Neel}
\hat{\rho}_{\rm Neel}
=
\bigotimes_{\mu\in\cal A}
\bigotimes_{\nu\in\cal B}
\hat{n}_\mu^\downarrow\,
\hat{\bar{n}}_\mu^\uparrow\,
\hat{n}_\nu^\uparrow\,
\hat{\bar{n}}_\nu^\downarrow
\,,
\end{eqnarray}
which is just the state with exactly one particle per site, but in 
alternating spin states, i.e., $s=\downarrow$ for $\mu\in\cal A$ and 
$s=\uparrow$ for $\nu\in\cal B$. 
Note that $\hat{n}_\mu^\downarrow$ is the projector on the 
$\ket{1}_\mu^\downarrow$ state 
$\hat{n}_\mu^\downarrow=\ket{1^\downarrow}_\mu\bra{1^\downarrow}$
while $\hat{\bar{n}}_\mu^\uparrow$
projects on the $\ket{0}_\mu^\uparrow$ state etc. 
As usual, this state (\ref{Neel}) breaks the original symmetry group of 
the Hamiltonian (\ref{Fermi-Hubbard-Hamiltonian}) containing particle-hole 
symmetry, $SU(2)$ invariance, and translational symmetry, down to a 
sub-group, which includes invariance under a combined spin-flip and 
particle-hole exchange etc. 

Let us stress that the N\'eel state (\ref{Neel}) is only the lowest-order 
approximation of the real ground state of the Heisenberg model 
(\ref{Heisenberg-Hamiltonian}), there are quantum spin fluctuations
of order $\ord(1/Z)$.  
These quantum spin fluctuations do not vanish in the limit $J\to 0$ since
$J$ only appears in the overall pre-factor in front of the 
Heisenberg Hamiltonian (\ref{Heisenberg-Hamiltonian}) while the internal 
structure remains the same. 
Only after adding a suitable staggered magnetic field 
(see \ref{groundcorr}), the N\'eel state (\ref{Neel}) 
is the exact unique ground state (for $J\to 0$). 
Either way, in analogy to the bosonic case, we can now use this fully 
factorizing state (\ref{Neel}) as the starting point for our $1/Z$-expansion. 

\section{Mott-N\'eel state}
\label{chargemodes}

Starting with the N\'eel state (\ref{Neel}) as the zeroth order in $1/Z$,
let us now derive the first-order corrections. 
To this end, let us consider the Heisenberg equations of motion 
\begin{eqnarray}
i\partial_t \hat{c}_{\mu s}
&=&
-\frac{J}{Z}\sum_{\kappa\neq \mu} T_{\mu\kappa}\hat{c}_{\kappa s}
+U\hat{c}_{\mu s} \hat{n}_{\mu \bar{s}}
\label{annihilation-operator}\\
i\partial_t \hat{c}_{\mu s}^\dagger
&=&
+\frac{J}{Z}\sum_{\kappa\neq \mu} T_{\mu\kappa}\hat{c}^\dagger_{\kappa s}
-U\hat{c}^\dagger_{\mu s} \hat{n}_{\mu \bar{s}}
\label{creation-operator}\\
i\partial_t\hat{n}_{\mu s}
&=&
\frac{J}{Z}\sum_{\kappa\neq \mu}T_{\mu\kappa}
\left(
\hat{c}_{\kappa s}^\dagger \hat{c}_{\mu s}-
\hat{c}_{\mu s}^\dagger \hat{c}_{\kappa s}
\right)
\nn
&=&
-i\partial_t\hat{\bar{n}}_{\mu s}
\label{number-operator}
\,,
\end{eqnarray}
where $\bar s$ denotes the spin label opposite to $s$, i.e., either 
$(s,\bar{s})=(\uparrow,\downarrow)$ or $(s,\bar{s})=(\downarrow,\uparrow)$.
If we now insert these evolution equations into the correlation functions
$\langle\hat{c}_{\mu a}^\dagger\hat{c}_{\nu b}
\hat{n}_{\mu\bar{a}}\hat{n}_{\nu\bar{b}}\rangle$, 
$\langle\hat{c}_{\mu a}^\dagger\hat{c}_{\nu b}
\hat{\bar n}_{\mu\bar{a}}\hat{n}_{\nu\bar{b}}\rangle$, 
$\langle\hat{c}_{\mu a}^\dagger\hat{c}_{\nu b}
\hat{n}_{\mu\bar{a}}\hat{\bar n}_{\nu\bar{b}}\rangle$, 
and 
$\langle\hat{c}_{\mu a}^\dagger\hat{c}_{\nu b}
\hat{\bar n}_{\mu\bar{a}}\hat{\bar n}_{\nu\bar{b}}\rangle$, 
we find that they form a closed set of equations to first order in $1/Z$, 
where we can neglect three-point correlations 
\begin{eqnarray}
i\partial_t
\langle \hat{c}_{\mu a}^\dagger\hat{c}_{\nu b}
\hat{n}_{\mu\bar{a}}\hat{n}_{\nu\bar{b}}\rangle
=
+\frac{J}{Z}T_{\mu\nu}
\langle \hat{c}_{\nu a}^\dagger\hat{c}_{\nu b}
\hat{n}_{\mu\bar{a}}\hat{n}_{\nu\bar{b}}\rangle_0
\nn
+\frac{J}{Z}\langle\hat{n}_{\mu \bar{a}}\rangle_0
\sum_{\kappa\neq \mu,\nu}T_{\mu\kappa}
\langle\hat{c}_{\kappa a}^\dagger\hat{c}_{\nu b}
(\hat{n}_{\kappa\bar{a}}+\hat{\bar{n}}_{\kappa\bar{a}})
\hat{n}_{\nu\bar{b}}\rangle
\nn
-\frac{J}{Z}\langle\hat{n}_{\nu \bar{b}}\rangle_0
\sum_{\kappa\neq \mu,\nu}T_{\nu\kappa}
\langle\hat{c}_{\mu a}^\dagger\hat{c}_{\kappa b}
\hat{n}_{\mu\bar{a}}(\hat{n}_{\kappa\bar{b}}+\hat{\bar{n}}_{\kappa\bar{b}})
\rangle
\nn
-\frac{J}{Z}T_{\mu\nu}
\langle \hat{c}_{\mu a}^\dagger\hat{c}_{\mu b}
\hat{n}_{\mu\bar{a}}\hat{n}_{\nu\bar{b}}\rangle_0
+\ord(1/Z^2)
\label{corr1}
\,,
\end{eqnarray}
where the expectation values $\langle\hat{n}_{\mu \bar{a}}\rangle_0$ and 
$\langle\hat{n}_{\nu \bar{b}}\rangle_0$ as well as those in the last line 
are taken in the zeroth-order N\'eel state (\ref{Neel}).
In complete analogy, we obtain for the remaining three correlators 
\begin{eqnarray}
i\partial_t 
\langle \hat{c}_{\mu a}^\dagger\hat{c}_{\nu b}
\hat{n}_{\mu\bar{a}}\hat{\bar{n}}_{\nu\bar{b}}\rangle
=
+\frac{J}{Z}T_{\mu\nu}
\langle \hat{c}_{\nu a}^\dagger\hat{c}_{\nu b}
\hat{n}_{\mu\bar{a}}\hat{\bar{n}}_{\nu\bar{b}}\rangle_0
\nn
+\frac{J}{Z}\langle\hat{n}_{\mu \bar{a}}\rangle_0
\sum_{\kappa\neq \mu,\nu}T_{\mu\kappa}
\langle\hat{c}_{\kappa a}^\dagger\hat{c}_{\nu b}
(\hat{n}_{\kappa\bar{a}}+\hat{\bar{n}}_{\kappa\bar{a}})
\hat{\bar{n}}_{\nu\bar{b}}\rangle
\nonumber\\
-\frac{J}{Z}\langle\hat{\bar{n}}_{\nu \bar{b}}\rangle_0
\sum_{\kappa\neq \mu,\nu}T_{\nu\kappa}
\langle\hat{c}_{\mu a}^\dagger\hat{c}_{\kappa b}
\hat{n}_{\mu\bar{a}}(\hat{n}_{\kappa\bar{b}}+\hat{\bar{n}}_{\kappa\bar{b}})
\rangle
\nonumber\\
-\frac{J}{Z}T_{\mu\nu}
\langle \hat{c}_{\mu a}^\dagger\hat{c}_{\mu b}
\hat{n}_{\mu\bar{a}}\hat{\bar{n}}_{\nu\bar{b}}\rangle_0
\nn
-U
\langle \hat{c}_{\mu a}^\dagger\hat{c}_{\nu b}
\hat{n}_{\mu \bar{a}}\hat{\bar{n}}_{\nu \bar{b}}\rangle
+\ord(1/Z^2)
\,,
\end{eqnarray}
as well as 
\begin{eqnarray}
i\partial_t 
\langle \hat{c}_{\mu a}^\dagger\hat{c}_{\nu b}
\hat{\bar{n}}_{\mu\bar{a}}\hat{n}_{\nu\bar{b}}\rangle
=
+\frac{J}{Z}T_{\mu\nu}
\langle \hat{c}_{\nu a}^\dagger\hat{c}_{\nu b}
\hat{\bar{n}}_{\mu\bar{a}}\hat{n}_{\nu\bar{b}}\rangle_0
\nn
+\frac{J}{Z}\langle\hat{\bar{n}}_{\mu \bar{a}}\rangle_0
\sum_{\kappa\neq \mu,\nu}T_{\mu\kappa}
\langle\hat{c}_{\kappa a}^\dagger\hat{c}_{\nu b}
(\hat{n}_{\kappa\bar{a}}+\hat{\bar{n}}_{\kappa\bar{a}})
\hat{n}_{\nu\bar{b}}\rangle
\nonumber\\
-\frac{J}{Z}\langle\hat{n}_{\nu \bar{b}}\rangle_0
\sum_{\kappa\neq \mu,\nu}T_{\nu\kappa}
\langle\hat{c}_{\mu a}^\dagger\hat{c}_{\kappa b}
\hat{\bar{n}}_{\mu\bar{a}}
(\hat{n}_{\kappa\bar{b}}+\hat{\bar{n}}_{\kappa\bar{b}})\rangle
\nonumber\\
-\frac{J}{Z}T_{\mu\nu}
\langle \hat{c}_{\mu a}^\dagger\hat{c}_{\mu b}
\hat{\bar{n}}_{\mu\bar{a}}\hat{n}_{\nu\bar{b}}\rangle_0
\nonumber\\
+U
\langle \hat{c}_{\mu a}^\dagger\hat{c}_{\nu b}
\hat{\bar{n}}_{\mu \bar{a}}\hat{n}_{\nu \bar{b}}\rangle
+\ord(1/Z^2)
\,,
\end{eqnarray}
and finally 
\begin{eqnarray}
i\partial_t 
\langle \hat{c}_{\mu a}^\dagger\hat{c}_{\nu b}
\hat{\bar{n}}_{\mu\bar{a}}\hat{\bar{n}}_{\nu\bar{b}}\rangle
=
+\frac{J}{Z}T_{\mu\nu}
\langle \hat{c}_{\nu a}^\dagger\hat{c}_{\nu b}
\hat{\bar{n}}_{\mu\bar{a}}\hat{\bar{n}}_{\nu\bar{b}}\rangle_0
\nn
+\frac{J}{Z}\langle\hat{\bar{n}}_{\mu \bar{a}}\rangle_0
\sum_{\kappa\neq \mu,\nu}T_{\mu\kappa}
\langle\hat{c}_{\kappa a}^\dagger\hat{c}_{\nu b}
(\hat{n}_{\kappa\bar{a}}+\hat{\bar{n}}_{\kappa\bar{a}})
\hat{\bar{n}}_{\nu\bar{b}}\rangle
\nonumber\\
-\frac{J}{Z}\langle\hat{\bar{n}}_{\nu \bar{b}}\rangle_0
\sum_{\kappa\neq \mu,\nu}T_{\nu\kappa}
\langle\hat{c}_{\mu a}^\dagger\hat{c}_{\kappa b}
\hat{\bar{n}}_{\mu\bar{a}}
(\hat{n}_{\kappa\bar{b}}+\hat{\bar{n}}_{\kappa\bar{b}})\rangle
\nonumber\\
-\frac{J}{Z}T_{\mu\nu}
\langle \hat{c}_{\mu a}^\dagger\hat{c}_{\mu b}
\hat{\bar{n}}_{\mu\bar{a}}\hat{\bar{n}}_{\nu\bar{b}}\rangle_0
+\ord(1/Z^2)
\label{sectors}
\,.
\end{eqnarray}
We observe that the spin structure is conserved in these equations, 
i.e., the four correlators containing 
$\hat{c}_{\mu\uparrow}^\dagger\hat{c}_{\nu\uparrow}$ decouple from 
those with 
$\hat{c}_{\mu\uparrow}^\dagger\hat{c}_{\nu\downarrow}$ etc. 
Thus we can treat the four sectors separately.  
Let us focus on the correlators containing 
$\hat{c}_{\mu\downarrow}^\dagger\hat{c}_{\nu\downarrow}$ and introduce
the following short-hand notation:
If ${\mu}\in\cal A$ and ${\nu}\in\cal B$, we denote the correlations by 
$\langle \hat{c}_{\mu \downarrow}^\dagger\hat{c}_{\nu \downarrow}
\hat{n}_{\mu\uparrow}\hat{n}_{\nu\uparrow}\rangle=f_{\mu\nu}^{1_A1_B}$,
and 
$\langle \hat{c}_{\mu \downarrow}^\dagger\hat{c}_{\nu \downarrow}
\hat{\bar{n}}_{\mu\uparrow}\hat{n}_{\nu\uparrow}\rangle=f_{\mu\nu}^{0_A1_B}$,
etc.
Inserting the zeroth-order N\'eel state (\ref{Neel}), 
we find four trivial equations which fully decouple
\begin{eqnarray}
i\partial_t f^{1_A0_B}_{\mu\nu}
&=&
-U f^{1_A0_B}_{\mu\nu}
\,,
\nn
i\partial_t f^{0_B1_A}_{\mu\nu}
&=&
+U f^{0_B1_A}_{\mu\nu}
\,,
\nn
i\partial_t f^{0_B0_B}_{\mu\nu}
&=&0
\,,
\nn
i\partial_t f^{1_A1_A}_{\mu\nu}
&=&0
\label{hom1}
\,.
\end{eqnarray}
Thus, if these correlations vanish initially, they remain zero 
(to first order in $1/Z$).
Setting these correlations (\ref{hom1}) to zero, 
we get four pairs of coupled equations 
\begin{eqnarray}
i\partial_t f^{0_A0_B}_{\mu\nu}
&=&
+\frac{J}{Z}\sum_{\kappa\neq{\mu,\nu}}T_{\mu\kappa}
f^{1_B0_B}_{\kappa\nu}
\,,
\nn
i\partial_t f^{1_B0_B}_{\mu\nu}
&=&
+\frac{J}{Z}\sum_{\kappa\neq{\mu,\nu}}T_{\mu\kappa}
f^{0_A0_B}_{\kappa\nu}
-Uf^{1_B0_B}_{\mu\nu}
\,,
\label{hom2}\\
i\partial_t f^{0_B0_A}_{\mu\nu}
&=&
-\frac{J}{Z}\sum_{\kappa\neq{\mu,\nu}}T_{\kappa\nu}
f^{0_B1_B}_{\mu\kappa}
\nn
i\partial_t f^{0_B1_B}_{\mu\nu}
&=&
-\frac{J}{Z}\sum_{\kappa\neq{\mu,\nu}}T_{\kappa\nu}
f^{0_B0_A}_{\mu\kappa}
+Uf^{0_B1_B}_{\mu\nu}
\,,
\\
i\partial_t f^{1_B1_A}_{\mu\nu}
&=&
+\frac{J}{Z}\sum_{\kappa\neq{\mu,\nu}}T_{\mu\kappa}
f^{0_A1_A}_{\kappa\nu}
\nn
i\partial_t f^{0_A1_A}_{\mu\nu}
&=&
+\frac{J}{Z}\sum_{\kappa\neq{\mu,\nu}}T_{\mu\kappa}
f^{1_B1_A}_{\kappa\nu}
+Uf^{0_A1_A}_{\mu\nu}
\,,
\\
i\partial_t f^{1_A1_B}_{\mu\nu}
&=&
-\frac{J}{Z}\sum_{\kappa\neq{\mu,\nu}}T_{\kappa\nu}
f^{1_A0_A}_{\mu\kappa}
\nn
i\partial_t f^{1_A0_A}_{\mu\nu}
&=&
-\frac{J}{Z}\sum_{\kappa\neq{\mu,\nu}}T_{\kappa\nu}
f^{1_A1_B}_{\mu\kappa}
-Uf^{1_A0_A}_{\mu\nu}
\,.
\label{hom12}
\end{eqnarray}
Again, since these equations do not have any non-vanishing source terms 
(to first order in $1/Z$), they can be set to zero if we start in an 
initially uncorrelated state. 
Note that they would acquire non-zero source terms if we go away from 
half-filling. 
The positive and negative eigenfrequencies of these modes behave as 
\bea\label{eigenmodes}
\omega_\mathbf{k}^\pm=\frac{U\pm\sqrt{U^2+4 J^2T_\mathbf{k}^2}}{2}
\,.
\ea
Thus we have soft modes which scale as 
$\omega_\mathbf{k}^-\sim J^2/U$ for small $J$ and hard modes 
$\omega_\mathbf{k}^+\approx U$.
These modes are important for making contact to the $t$-$J$ 
model~\cite{A94} which describes the low-energy excitations of the 
Fermi-Hubbard Hamiltonian~(\ref{Fermi-Hubbard-Hamiltonian}) for small $J$
away from half-filling.
However, at half-filling, we can set them to zero. 
After doing this, we are left with four coupled equations, which do have 
non-vanishing source terms 
\begin{eqnarray}
i\partial_t f^{0_A0_A}_{\mu\nu}
&=&
\frac{J}{Z}\sum_{\kappa\neq{\mu,\nu}}
\left\{
T_{\mu\kappa}
f^{1_B0_A}_{\kappa\nu}
-T_{\kappa\nu}
f^{0_A1_B}_{\mu\kappa}
\right\}
\label{charge1}
\,,
\\
i\partial_t f^{0_A1_B}_{\mu\nu}
&=&
\frac{J}{Z}\sum_{\kappa\neq{\mu,\nu}}
\left\{
T_{\mu\kappa}
f^{1_B1_B}_{\kappa\nu}
-T_{\kappa\nu}
f^{0_A0_A}_{\mu\kappa}
\right\}
\nonumber\\
& &
+U f^{0_A1_B}_{\mu\nu}-\frac{J}{Z}T_{\mu\nu}
\,,
\\
i\partial_t f^{1_B0_A}_{\mu\nu}
&=&
\frac{J}{Z}\sum_{\kappa\neq{\mu,\nu}}
\left\{
T_{\mu\kappa}
f^{0_A0_A}_{\kappa\nu}
-T_{\kappa\nu}
f^{1_B1_B}_{\mu\kappa}
\right\}
\nonumber\\
& &
-U f^{1_B0_A}_{\mu\nu}+\frac{J}{Z}T_{\mu\nu}
\,,
\\
i\partial_t f^{1_B1_B}_{\mu\nu}
&=&
\frac{J}{Z}\sum_{\kappa\neq{\mu,\nu}}
\left\{
T_{\mu\kappa}
f^{0_A1_B}_{\kappa\nu}
-T_{\kappa\nu}
f^{1_B0_A}_{\mu\kappa}
\right\}
\,.
\label{charge4}
\end{eqnarray}
Due to the source terms $JT_{\mu\nu}/Z$, these modes will develop 
correlations if we slowly (or suddenly) switch on the hopping rate $J$,
even if there are no correlations initially. 
The eigenfrequencies of these (charge) modes behave as 
\begin{eqnarray}
\label{omega-fermi}
\omega_\mathbf{k}=\sqrt{U^2+4 J^2T_\mathbf{k}^2}
\,.
\end{eqnarray}
A similar dispersion relation can be derived from a mean-field approach~\cite{F91}. 
In contrast to the bosonic case, the origin of the Brillouin zone at 
$\mathbf{k}=0$ does not have minimum but actually maximum excitation 
energy $\omega_\mathbf{k}$.
The minimum is not a point but a hyper-surface where $T_\mathbf{k}=0$
(or, more generally, $T_\mathbf{k}^2$ assumes its minimum).
After Fourier transformation of (\ref{charge1})-(\ref{charge4}) we find 
that the equations of motion conserve a bilinear quantity, that is
\begin{eqnarray}\label{invfermi}
\partial_t\left[
\left(f^{1_B1_B}_\mathbf{k}-1\right)f^{1_B1_B}_\mathbf{k}+
f^{0_A1_B}_\mathbf{k}f^{1_B0_A}_\mathbf{k}
\right]=0\,.
\end{eqnarray}
This relation holds, as in the bosonic case, also for time-dependent $J(t)$.


\subsection{Ground-state correlations}
\label{groundcorr}

In complete analogy to the bosonic case, we now imagine switching $J$ 
adiabatically from zero (where all the charge fluctuations vanish) to 
a finite value. 
In order to operate this adiabatic switching, we must start in principle at  
$J=0$ from a non degenerate ground state. 
This is accomplished by adding a term into the fermion Hamiltonian:
\begin{eqnarray}
\hat H
\rightarrow 
\hat H -
\sum_{\mu}
(A_{\mu\downarrow}\hat{n}_\mu^\downarrow +
A_{\mu\uparrow} \hat{n}_\mu^\uparrow)
\,.
\end{eqnarray}
If we choose the magnetic field as 
$A_{\mu\downarrow}(x_\mu\in \mathcal{A})=a$,
$A_{\mu\downarrow}(x_\mu\in \mathcal{B})=
A_{\mu\uparrow}(x_\mu\in \mathcal{A})=0$,  
and 
$A_{\mu\uparrow}(x_\mu\in \mathcal{B})=a$,
the N\'eel state is the unique ground state for $J=0$ at half filling. 
Repeating the steps in Eqs.~(\ref{charge1}-\ref{charge4}) and 
(\ref{hom2}-\ref{hom12}) by including this term,
the eigenfrequencies (\ref{eigenmodes}) and (\ref{omega-fermi}) read now
\begin{eqnarray}
\label{eigenmodesstaggerd}
\omega_\mathbf{k}^\pm=\frac{U+a\pm\sqrt{4J^2T_\mathbf{k}^2+(U-a)^2}}{2}\,,
\end{eqnarray}
and 
\begin{eqnarray}
\omega_\mathbf{k}=\sqrt{4J^2T_\mathbf{k}^2+(U-a)^2}\,.
\end{eqnarray}
After adiabatic switching,  we find in the limit $a=0$ 
the following non-zero ground-state correlations 
\begin{eqnarray}
\label{ground-11}
f^{1_B1_B}_{\mu\nu,\mathrm{ground}}
&=&
-f^{0_A0_A}_{\mu\nu,\mathrm{ground}}
\\
&=&
\frac{1}{2N}\sum_{\mathbf{k}}
\left(1-\frac{U}{\omega_\mathbf{k}}\right)
e^{i(\mathbf{x}_\mu-\mathbf{x}_\nu)\cdot\mathbf{k}}
\,,
\nonumber
\\
f^{1_B0_A}_{\mu\nu,\mathrm{ground}}
&=&
{f}^{0_A1_B}_{\mu\nu,\mathrm{ground}}
=
\frac{1}{N}\sum_{\mathbf{k}}
\frac{JT_\mathbf{k}}{\omega_\mathbf{k}}\,
e^{i(\mathbf{x}_\mu-\mathbf{x}_\nu)\cdot\mathbf{k}}
\,,
\label{ground-10}
\end{eqnarray}
which reproduce the expressions obtained in Ref.~\cite{LPM69}.
Somewhat similar to the Bose-Hubbard model, the symmetric combination 
(\ref{ground-11}) scales with $J^2$ for small $J$ while the other 
(\ref{ground-10}) starts linearly in $J$. 
Other correlators such as 
$\langle\hat{c}_{\mu\downarrow}^\dagger\hat{c}_{\nu\downarrow}\rangle$
can be obtained from these expressions.
For example, if $\mu$ and $\nu$ are in $\cal A$, we find, 
using $\hat{n}_{\mu\uparrow}+\hat{\bar n}_{\mu\uparrow}=1$ 
and $\hat{n}_{\nu\uparrow}+\hat{\bar n}_{\nu\uparrow}=1$
\bea
\langle\hat{c}_{\mu\downarrow}^\dagger\hat{c}_{\nu\downarrow}\rangle
&=&
f_{\mu\nu}^{1_A1_A}
+f_{\mu\nu}^{0_A1_A}
+f_{\mu\nu}^{1_A0_A}
+f_{\mu\nu}^{0_A0_A}
\nn
&=&
f_{\mu\nu}^{0_A0_A}
\,.
\ea

\subsection{Quantum depletion}

To zeroth order, i.e., in the N\'eel state (\ref{Neel}), we have 
$\langle\hat{n}_{\mu\uparrow}\hat{n}_{\mu\downarrow}\rangle=0$.
Thus this quantity 
$\langle\hat{n}_{\mu\uparrow}\hat{n}_{\mu\downarrow}\rangle$
measures the deviation from this zeroth-order N\'eel state (\ref{Neel})
due to quantum charge fluctuations.
In order to calculate 
$\langle\hat{n}_{\mu\uparrow}\hat{n}_{\mu\downarrow}\rangle$, 
we also need some of the other sectors discussed after (\ref{sectors}).
Obviously, the correlators containing 
$\hat{c}_{\mu\uparrow}^\dagger\hat{c}_{\nu\uparrow}$ 
behave in the same way as those with 
$\hat{c}_{\mu\downarrow}^\dagger\hat{c}_{\nu\downarrow}$
after interchanging the sub-lattices $\cal A$ and $\cal B$.
Thus a completely analogous system of differential equations exists for 
the correlations of the form 
$\langle \hat{c}_{\mu \uparrow}^\dagger\hat{c}_{\mu \uparrow}
\hat{n}_{\mu\downarrow}\hat{n}_{\nu\downarrow}\rangle
=g_{\mu\nu}^{1_A1_B}$ etc. 
If we insert (\ref{number-operator}) in order to calculate 
$i\partial_t\langle\hat{n}_{\mu\uparrow}\hat{n}_{\mu\downarrow}\rangle$, 
we find that these two sectors are enough for deriving 
$\langle\hat{n}_{\mu\uparrow}\hat{n}_{\mu\downarrow}\rangle$. 
Assuming $\mu\in\cal B$ for simplicity, we find  
\begin{eqnarray}
i\partial_t\langle \hat{n}_{\mu s}\hat{n}_{\mu \bar{s}}\rangle
=
-\frac{J}{Z}\sum_{\kappa\neq\mu}T_{\kappa\mu}
\Big\{
g_{\mu\kappa}^{1_B1_A}+g_{\mu\kappa}^{1_B0_A}+f_{\mu\kappa}^{1_B1_A}
\nn
+f_{\mu\kappa}^{1_B0_A}-g_{\kappa\mu}^{1_A1_B}-g_{\kappa\mu}^{0_A1_B}
-f_{\kappa\mu}^{1_A1_B}-f_{\kappa\mu}^{0_A1_B}
\Big\}
\,.
\end{eqnarray}
Setting the correlations with vanishing source terms to zero, we get 
\begin{eqnarray}
i\partial_t\langle \hat{n}_{\mu s}\hat{n}_{\mu \bar{s}}\rangle
&=&
-\frac{J}{Z}\sum_{\kappa\neq\mu}T_{\kappa\mu}
\Big\{
f_{\mu\kappa}^{1_B0_A}
-f_{\kappa\mu}^{0_A1_B}
\Big\}\nonumber\\
& =&-\frac{1}{N}\sum_\mathbf{k}JT_\mathbf{k}\Big\{
f_{\mathbf{k}}^{1_B0_A}
-f_{\mathbf{k}}^{0_A1_B}
\Big\}
\nn
&=&
\frac{i}{N}\sum_\mathbf{k}\partial_t f_{\mathbf{k}}^{1_B1_B}
\,.
\end{eqnarray}
Thus, in the ground state, the quantum depletion reads 
\begin{eqnarray}
\label{fermi-depletion}
\langle\hat{n}_{\mu s}\hat{n}_{\mu \bar{s}}\rangle
=
\langle\hat{\bar{n}}_{\mu s}\hat{\bar{n}}_{\mu \bar{s}}\rangle
=
\frac{1}{2N}\sum_\mathbf{k}
\left(1-\frac{U}{\omega_\mathbf{k}}\right)
\,.
\end{eqnarray}
As one would expect, this quantity scales with $J^2$ for small $J$. 
The results (\ref{ground-11}), (\ref{ground-10}), and (\ref{fermi-depletion})
can also be obtained via other approaches, such as the spin density wave 
ansatz \cite{BEM2009} (which is related to dynamical mean field theory 
according to Ref.~\cite{GKKR96}). 

\subsection{Spin modes}

So far, we have considered expectations values such as 
$\langle\hat{c}_{\mu a}^\dagger\hat{c}_{\nu b}
\hat{n}_{\mu\bar{a}}\hat{n}_{\nu\bar{b}}\rangle$, 
where -- apart from the number operators $\hat{n}_{\mu\bar{a}}$ and 
$\hat{n}_{\nu\bar{b}}$ -- one particle is annihilated at site $\nu$ 
and one is created at site $\mu$. 
These operator combinations correspond to a change of the occupation 
numbers and are thus called charge modes.
However, as already indicated in Section~\ref{Fermi-Hubbard Model}, 
there are also other modes which leave the total occupation number 
of all lattice sites unchanged. 
Examples are 
$\langle\hat{c}_{\mu s}^\dagger\hat{c}_{\mu\bar{s}}
\hat{c}_{\nu\bar{s}}^\dagger\hat{c}_{\nu s}\rangle$
or 
$\langle\hat{n}_{\mu a}\hat{n}_{\nu b}\rangle$
or combinations thereof. 
Many of these combinations can be expressed in terms of the 
effective spin operators in (\ref{spin-operators}) via 
$\langle\hat{S}_{\mu}^i\hat{S}_{\nu}^j\rangle$.
As one would expect from our study of the Bose-Hubbard model, 
the evolution of these spin modes vanishes to first order in $1/Z$ 
\bea
\partial_t\langle\hat{S}_{\mu}^i\hat{S}_{\nu}^j\rangle=\ord(1/Z^2)
\,,
\ea
consistent with the Heisenberg Hamiltonian (\ref{Heisenberg-Hamiltonian}).
In analogy to the $\langle\hat{n}_{\mu}\hat{n}_{\nu}\rangle$-correlator 
in the bosonic case, one has to go to second order $\ord(1/Z^2)$ in order
to calculate these quantities.
Fortunately, the charge modes discussed above do not couple to these spin 
modes to first order in $1/Z$ and hence we can omit them to this level
of accuracy. 

\section{Quench dynamics}
\label{fermi-noneq}

\begin{center}
\begin{figure}[h]
\includegraphics{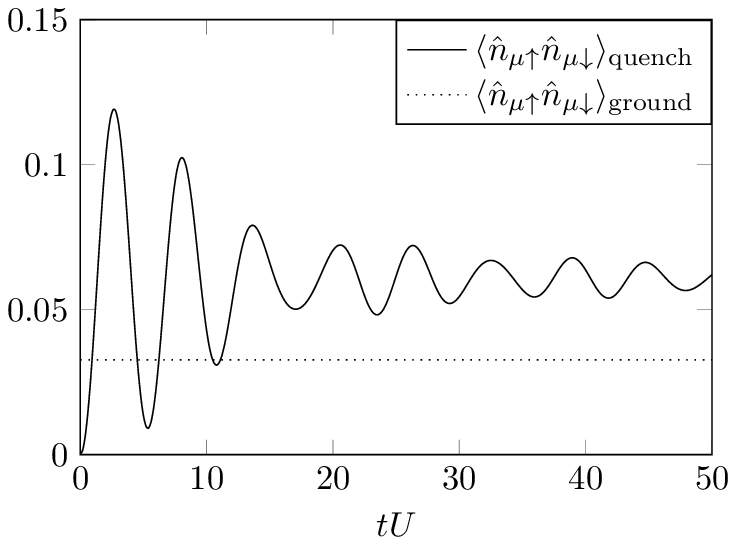}
\includegraphics{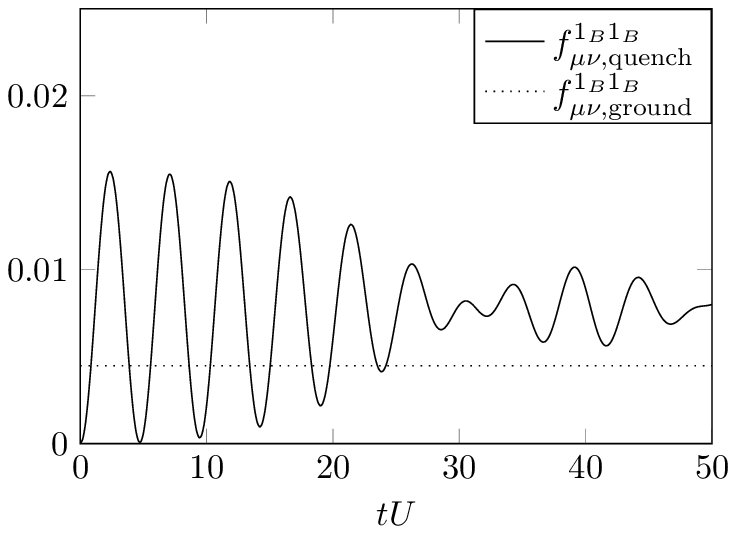}
\begin{center}
\includegraphics{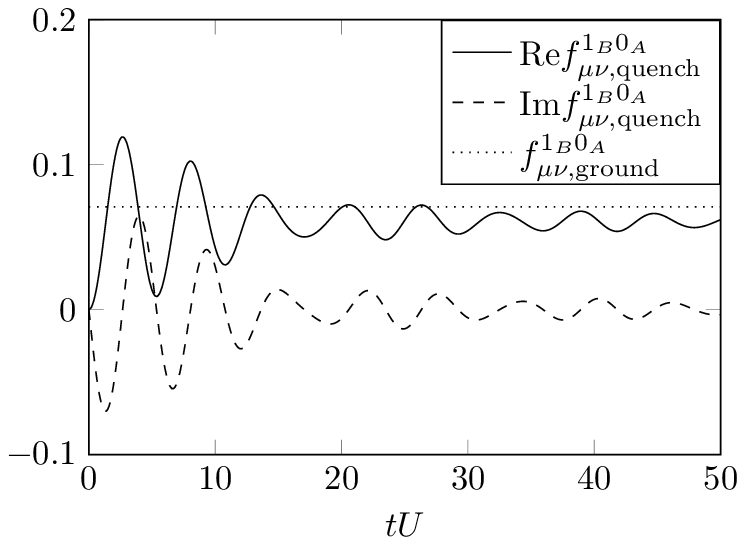}
\end{center}
\caption{Time-dependence of the quantum depletion, the nearest-neighbor 
correlation function $f^{1_B0_A}_{\mu\nu}$, and the next-to-nearest-neighbor 
correlation function $f^{1_B1_B}_{\mu\nu}$ in three dimensions 
after a quench within the Mott phase according to $J/U=0\to 0.5$ 
in comparison to their ground-state values.}
\label{quenchfermi}
\end{figure}
\end{center}

Now we consider a quantum quench, i.e., a sudden switch from $J=0$ 
to some finite value of $J$.
To this end, we start with the Mott-N\'eel state~(\ref{Neel}), 
which is an exact eigenstate of the Hamiltonian for $J=0$, 
and solve the first-order (in $1/Z$) equations for the correlations.
This provides a good approximation at least for short and intermediate 
times, before $1/Z^2$-corrections (such as the soft spin modes) 
start to play a role.  
Following this strategy, we find the following non-vanishing correlations
\begin{eqnarray}
f^{1_B1_B}_{\mu\nu,\mathrm{quench}}
&=&
-f^{0_A0_A}_{\mu\nu,\mathrm{quench}}
\\
&=&
\frac{1}{N}\sum_{\mathbf{k}}
2 J^2 T_{\mathbf{k}}^2\,
\frac{1-\cos(\omega_\mathbf{k} t)}{\omega_\mathbf{k}^2}\,
e^{i(\mathbf{x}_\mu-\mathbf{x}_\nu)\cdot\mathbf{k}}
\,,
\nonumber
\end{eqnarray}
and 
\begin{eqnarray}
f^{1_B0_A}_{\mu\nu,\mathrm{quench}}
&=&
\left({f}^{0_A1_B}_{\mu\nu,\mathrm{quench}}\right)^*
\\
&=&
\frac{1}{N}\sum_{\mathbf{k}}
J T_\mathbf{k}U
\frac{1 - \cos(\omega_\mathbf{k} t)}{\omega_\mathbf{k}^{2}}
e^{i(\mathbf{x}_\mu-\mathbf{x}_\nu)\cdot\mathbf{k}}
\nn
& &
-\frac{i}{N}\sum_{\mathbf{k}}
J T_\mathbf{k}\,
\frac{\sin(\omega_\mathbf{k} t)}{\omega_\mathbf{k}}\,
e^{i(\mathbf{x}_\mu-\mathbf{x}_\nu)\cdot\mathbf{k}}
\nonumber
\,.
\end{eqnarray}
Again, these correlations equilibrate to a quasi-stationary value,
which is, however, not thermal.  
For some of these correlations, this quasi-stationary value lies 
even {\em below} the ground-state correlation, see Fig.~\ref{quenchfermi}. 
The probability to have two or zero particles at a site reads 
\begin{eqnarray}
\label{docc}
\langle\hat{n}_{\mu s}\hat{n}_{\mu \bar{s}}\rangle_\mathrm{quench}
&=&
\langle\hat{\bar{n}}_{\mu s}\hat{\bar{n}}_{\mu \bar{s}}\rangle_\mathrm{quench}
\\
&=&
\frac{1}{N}\sum_\mathbf{k}
2 J^2 T_{\mathbf{k}}^2\,
\frac{1-\cos(\omega_\mathbf{k} t)}{\omega_\mathbf{k}^2}
\,.
\nonumber
\end{eqnarray}
This quantity also equilibrates to a quasi-stationary value of order $1/Z$.
In analogy to the bosonic case, this quasi-stationary value could be explained 
by a small effective temperature -- but this small effective temperature then
does not work for the other observables, e.g., the correlations. 

The time-evolution of the quantum depletion in Fig.~\ref{quenchfermi} 
can be compared with the results of Ref.~\cite{KE08} where the (integrable) 
Fermi-Hubbard model in one dimension with long-range hopping 
(i.e., $T_{\mu\nu}$ contributes not just for nearest neighbors)
is considered and we observe qualitative agreement
(see, e.g., Fig.~1d in Ref.~\cite{KE08}). 
Unfortunately, a quantitative comparison of our results for the 
higher-dimensional Fermi-Hubbard model is impeded by the lack of data for 
the regime under consideration in our present work. 
For instance, the Fermi-Hubbard model in one and two dimensions is studied 
in Ref.~\cite{GA12}, but there a quench from $U>0$ to $U=0$ is considered.
As another example, the quench from $U=0$ to $U>0$ 
(but still in the metallic phase, i.e., for weak $U$)
is investigated in Ref.~\cite{MK08}, where three temporal regimes are identified: 
short times (build-up and oscillation of correlations), intermediate times 
(quasi-equilibration), and late times (thermalization). 
The first two temporal regimes can be recovered in complete analogy within 
our first-order $1/Z$-approach, but the late-time behavior (thermalization)
requires higher orders in $1/Z$. 

\section{Conclusions}
\label{Conclusions}

In summary, we studied the quantum dynamics of the Bose and Fermi Hubbard 
model after a quench within the Mott phase.
To this end, we employed a formal expansion into powers of $1/Z$ based on 
the hierarchy of correlations. 
In comparison to other approaches (as mentioned in the Introduction, 
for example), this method facilitates an iterative approximate analytical 
solution for the time dependence of the reduced density matrices and their 
ground state values.
It is particularly suited for the strongly interacting regime in higher 
dimensions and can be applied to a general lattice structure $T_{\mu\nu}$
of arbitrary size.  
Since our method is based on an expansion into powers of the (small) 
control parameter $1/Z$, it provides a unique classification which effect 
occurs at which order in $1/Z$.
This fact is also related to the somewhat disadvantageous features of our 
approach, for example the fact that the correct treatment of the soft spin 
modes and the late-time dynamics requires higher orders in $1/Z$.
Furthermore, we cannot describe the transition between the Mott insulator 
and the metallic state in the Fermi-Hubbard model within our first-order 
approach. 

As one application, we derive the spread of correlations and obtain 
an effective light-cone structure (via the saddle-point approximation). 
Furthermore, we found that the considered observables settle down to a 
quasi-equilibrium state after some time -- but this state is not thermal. 
More precisely, the on-site density matrix settles down to a state which 
could be described by a thermal ensemble but the two-point correlations do
not fit this thermal state.

Thus, real thermalization -- if it occurs at all -- requires much longer 
times scales.
This seems to be a generic feature and has been discussed for 
bosonic~\cite{KLA07,CFMSE08,FCMSE08,CDEO08} and fermionic 
systems~\cite{U09,MK08,EKW10,MK09,MK10,SGJ10,EKW09,luttinger}
and is sometimes called ``pre-thermalization''~\cite{BBS03,BBW04}.
This phenomenon can be visualized via the following intuitive picture:
The excited state generated by the quench can be viewed as a highly 
coherent superposition of correlated quasi-particles.
During the subsequent quantum evolution, these quasi-particles 
disperse and randomize their relative phases -- which results in 
a quasi-stationary state. 
However, the quasi-particles still retain their initial spectrum
(in energy and quasi-momentum), which could be approximately described 
by a generalized Gibbs ensemble (i.e., a momentum-dependent temperature). 
In this picture, thermalization requires the exchange of energy and 
momentum between these quasi-particles due to multiple collisions, 
which changes the one-particle spectrum and takes much longer. 
Ergo, one would expect a separation of time scales -- i.e., first (quasi)
equilibration and only much later thermalization -- for many systems 
in condensed matter, where the above quasi-particle picture applies. 

Within our $1/Z$ approach, the interaction between the quasi-particles 
(responsible for the exchange of energy and momentum by multiple collisions) 
correspond to higher orders in $1/Z$.
Since they become relevant at time scales much longer than the initial 
dephasing time considered here, one would expect that it is possible to 
derive some sort of Boltzmann equation for these long time scales. 

\section*{Acknowledgments} 

The authors acknowledge valuable discussions with W.~Hofstetter, S.~Kehrein, 
C.~Kollath, A.~Rosch, M.~Vojta and many others.   
F.Q.\ is supported by the Templeton foundation (grant number JTF 36838). 
This work was supported by the DFG (SFB-TR12). 

\appendix
\section{Derivation of the hierarchy}
\label{hierarchyApp}

In this Appendix, we derive the hierarchical set of equations for the 
correlation functions.
The quantum evolution of the on-site density matrix can be derived
by tracing von Neumann's equation (\ref{Liouville})
over all lattice sites but $\mu$ and exploiting the invariance of the 
trace under cyclic permutations 
\begin{eqnarray}
\label{singlesite}
i\partial_t \hat{\rho}_\mu
&=&
\frac{1}{Z}\tr_{\not\mu}
\left\{\sum_{\alpha,\beta\neq\mu}\mathcal{L}_{\alpha\beta}\hat{\rho}
+\sum_{\alpha\neq \mu}\mathcal{L}^S_{\alpha\mu}\hat{\rho}\right\}
\nn
&&
+\tr_{\not\mu}\left\{\sum_{\alpha\neq\mu}
\mathcal{L}_\alpha\hat{\rho}+\mathcal{L}_\mu\hat{\rho}\right\}
\nonumber\\
&=&
\frac{1}{Z}\sum_{\alpha\neq \mu} 
\mathcal{L}^S_{\mu\alpha}\tr_{\alpha}\{\hat{\rho}_{\mu\kappa}\}
+\mathcal{L}_\mu\hat{\rho}_\mu\,.
\end{eqnarray}
Using the definition of the two-point correlations given in 
(\ref{correlated-parts}), we arrive at (\ref{one-site}).
Similarly, the differential equation for the two-particle density matrix can 
be deduced by tracing over all lattice sites but $\mu$ and $\nu$,
\begin{eqnarray}
i\partial_t \hat{\rho}_{\mu\nu}
&=&
i\left(\partial_t \hat{\rho}_{\mu\nu}^\mathrm{corr}+
\hat{\rho}_\mu \partial_t \hat{\rho}_\nu+
\hat{\rho}_\nu \partial_t \hat{\rho}_\mu\right)
\nonumber\\
&=&
\frac{1}{Z}\sum_{\alpha\neq \mu\nu}
\left(
\tr_\alpha\left\{\mathcal{L}_{\mu\kappa}^S\hat{\rho}_{\mu\nu\alpha}\right\}
+
\tr_\alpha\left\{\mathcal{L}_{\kappa\nu}^S\hat{\rho}_{\mu\nu\alpha}\right\}
\right)
\nonumber\\
&&
+\frac{1}{Z}\mathcal{L}_{\mu\nu}^S\hat{\rho}_{\mu\nu}
+\mathcal{L}_\mu \hat{\rho}_{\mu\nu}+\mathcal{L}_\nu \hat{\rho}_{\mu\nu}\,.
\end{eqnarray}
With the definitions (\ref{correlated-parts}) and the 
time-evolution for the single-site density matrix (\ref{singlesite}),
we find for the two-point correlation functions (\ref{two-sites}).
The equations (\ref{one-site}) and (\ref{two-sites})
preserve the hierarchy in time if initially
$\hat{\rho}_\mu=\mathcal{O}(Z^0)$ and 
$\hat{\rho}_{\mu\nu}^\mathrm{corr}=\mathcal{O}(1/Z)$ holds.
In order to derive the full hierarchy,
we define the generating functional
\bea
{\cal F}(\hat\alpha)
=
{\cal F}(\{\hat\alpha_\mu\}) 
=
\ln\left[
\tr\left\{\hat\rho\bigotimes_\mu(\mathbf{1}_\mu+\hat\alpha_\mu)\right\}
\right]
\,,
\eea 
where $\hat\rho$ is the density matrix of the full lattice and 
\begin{eqnarray}
\hat\alpha_\mu=\sum_{m,n}\alpha_\mu^{m,n}|m\rangle_\mu\langle n|
\end{eqnarray}
are arbitrary operators acting on the Hilbert spaces associated to the 
lattice sites $\mu$ with the local basis $\{\ket{n}_\mu\}$.
The role of this functional is to generate all correlated density matrices 
via the derivatives with respect to these operators $\hat\alpha_\mu$ which 
are defined via 
\begin{eqnarray}
\frac{\partial{\cal F(\{\alpha\})}}{{\partial\hat\alpha_\mu}}
&=&
\sum_{m,n}|n\rangle_\mu\langle m|\,
\frac{\partial \cal F(\{\alpha\})}{\partial \alpha^{m,n}_\mu}
\nn
&=&
\sum_{m,n}|n\rangle_\mu\langle m|\,
\frac{\partial{\cal F}(\{\alpha\})}
{\partial\, {_\mu\!\bra{m}}\hat\alpha_\mu\ket{n}_\mu}
\,.
\end{eqnarray}
If we consider an ensemble ${\cal S}=\{\mu_1, \dots ,\mu_\ell\}$  
of $\ell$ different lattice sites $\mu_1\not= \dots \not=\mu_\ell$, 
we obtain the correlation operators via 
\bea
\hat\rho^{\rm corr}_{\cal S} 
=
\left.
\frac{\partial}{\partial\hat\alpha_{\mu_1}}
\frac{\partial}{\partial\hat\alpha_{\mu_2}}
\dots 
\frac{\partial}{\partial\hat\alpha_{\mu_\ell}}
{\cal F}(\hat \alpha)\right|_{\hat \alpha=0} 
\,.
\eea
These operators are related to the corresponding reduced density matrix 
operator $\rho_{\cal S}$ through the relation 
\bea
\hat\rho_{\cal S} 
= 
\hat\rho_{\mu_1 \dots \mu_\ell}
=
\sum_{\cup_i{\cal P}_i={\cal S}}\prod_i\hat\rho^{\rm corr}_{{\cal P}_i}
\ea
where the sum runs over all possible segmentations of the subset ${\cal S}$
into partitions ${\cal P}_i$ starting from the whole subset 
${\cal P}={\cal S}$ and ranging to single lattice sites 
${\cal P}_i=\{\mu\}$ where 
$\hat\rho^{\rm corr}_{{\cal P}_i=\{\mu\}}=\hat\rho_\mu$ is understood. 
For two and three lattice sites, the above equation reproduces 
Eq.~(\ref{correlated-parts}). 

Our derivation is based on the following scaling hierarchy of 
correlations:
\bea
\label{hierarchy1}
\hat\rho_{\cal S}^{\rm c}=\ord\left(Z^{1-|\cal S|}\right)
\ea
where $|\cal S|$ is the number $\ell$ of lattice sites in the set $\cal S$.
From the Liouville equation (\ref{Liouville}), 
the temporal evolution of ${\cal F}$ is given by 
\begin{widetext}
\bea
\label{eqgf}
i \partial_t 
{\cal F}(\hat\alpha)
=
\sum_\mu\tr_\mu 
\left\{
\hat\alpha_\mu \li_\mu \frac{\partial{\cal F}}{\partial\hat\alpha_\mu}
\right\}
+
\frac{1}{Z} \sum_{\mu,\nu} \tr_{\mu\nu} 
\left\{
(\hat\alpha_\mu + \hat\alpha_\nu +\hat\alpha_\mu \hat\alpha_\nu) 
\li_{\mu \nu} 
\left(
\frac{\partial^2{\cal F}}{\partial\hat\alpha_\mu\partial\hat\alpha_\nu}
+
\frac{\partial{\cal F}}{\partial\hat\alpha_\mu}
\frac{\partial{\cal F}}{\partial\hat\alpha_\nu}
\right)
\right\}
\,.
\eea
By taking successive derivatives and using the generalized Leibniz rule 
\bea
\frac{\partial}{\partial\hat\alpha_{\mu_1}}
\frac{\partial}{\partial\hat\alpha_{\mu_2}}
\dots 
\frac{\partial}{\partial\hat\alpha_{\mu_\ell}}
\left[{\cal F}(\hat \alpha)\right]^2
=
\sum_{{\cal P}\subseteq {\cal S}}^{{\cal P}\cup\bar{\cal P}={\cal S}}  
\left[
\left(
\prod_{\mu_i \in {\cal P}}
\frac{\partial}{\partial\hat\alpha_{\mu_i}}
\right)
{\cal F}(\hat \alpha)
\right]
\left[
\left(
\prod_{\mu_j \in \bar{\cal P}}
\frac{\partial}{\partial\hat\alpha_{\mu_j}}
\right)
{\cal F}(\hat \alpha)
\right]
\,,
\eea
as well as the the property 
\bea
\frac{\partial^2{\cal F}(\hat \alpha)}{\partial\hat\alpha_{\mu}^2}
=
\frac{\partial}{\partial\hat \alpha_{\mu}}
\frac{\partial}{\partial\hat \alpha_{\mu}}
{\cal F}(\hat \alpha)
=
-
\frac{\partial{\cal F}(\hat \alpha)}{\partial\hat\alpha_{\mu}}
\frac{\partial{\cal F}(\hat \alpha)}{\partial\hat\alpha_{\mu}}
=
-
\left(\frac{\partial{\cal F}(\hat \alpha)}{\partial\hat\alpha_{\mu}}\right)^2
\,,
\ea
we establish the following set of 
equations for the correlated density matrices: 
\begin{eqnarray}
\label{general}
i \partial_t\hat\rho^{\rm corr}_{\cal S}
&=&
\frac{1}{Z}
\sum_{\mu,\nu\in{\cal S}} 
\sum_{{\cal P}\subseteq{\cal S}\setminus\{\mu,\nu\}}^{{\cal P}\cup\bar{\cal P}
={\cal S}\setminus\{\mu,\nu\}}
\Bigg\{
\li_{\mu \nu}\,
\hat\rho^{\rm corr}_{\{\mu\}\cup{\cal P}}\,
\hat\rho^{\rm corr}_{\{\nu\}\cup\bar{\cal P}}
-\tr_{\nu}\Bigg[
\li^S_{\mu \nu}(\hat\rho^{\rm corr}_{\{\mu,\nu\}\cup\bar{\cal P}} 
+
\sum_{{\cal Q}\subseteq\bar{\cal P}}
^{{\cal Q}\cup\bar{\cal Q}=\bar{\cal P}}
\hat\rho^{\rm corr}_{\{\mu\}\cup{\cal Q}}\,
\hat\rho^{\rm corr}_{\{\nu\}\cup\bar{\cal Q}}
)
\Bigg]
\hat\rho^{\rm corr}_{\{\nu\}\cup{\cal P}}
\Bigg\}
\nn
&&
+
\sum_{\mu \in {\cal S}} \li_\mu \hat\rho^{\rm corr}_{\cal S}
+
\frac{1}{Z}\sum_{\mu,\nu\in{\cal S}}
\li_{\mu \nu}\,\hat\rho^{\rm corr}_{\cal S} 
+ 
\frac{1}{Z}
\sum_{\kappa\notin{\cal S}} \sum_{\mu\in{\cal S}} 
\tr_{\kappa}
\Bigg[
\li^S_{\mu \kappa}
\hat\rho^{\rm corr}_{{\cal S}\cup {\kappa}}
+ 
\sum_{{\cal P}\subseteq{\cal S}\setminus\{\mu\}}^{{\cal P}\cup\bar{\cal P}
={\cal S}\setminus\{\mu\}}
\li^S_{\mu \kappa}
\hat\rho^{\rm corr}_{\{\mu\}\cup{\cal P}}\,
\hat\rho^{\rm corr}_{\{\kappa\}\cup\bar{\cal P}}
\Bigg]
\,.
\end{eqnarray}
For $\ell=1$ and $\ell=2$ we recover the equations 
(\ref{one-site}) and (\ref{two-sites}).
A careful inspection of this set of equations shows that the hierarchy 
in (\ref{hierarchy1}) is preserved in time: 
Imposing the scaling 
$\hat\rho^{\rm corr}_{\cal S}=\ord(Z^{1-|\cal S|})$ on the r.h.s.\ of the 
above equation, we find that the time derivative on the l.h.s.\ does also 
satisfy the hierarchy (\ref{hierarchy1}). 
Therefore, inserting (\ref{hierarchy1}) into (\ref{general}) and taking the 
limit $Z\to\infty$, we obtain the leading-order contributions
\bea
\label{general1}
i \partial_t\hat\rho^{\rm corr}_{\cal S}
&=&
\frac{1}{Z}
\sum_{\mu,\nu\in{\cal S}} 
\sum_{{\cal P}\subseteq{\cal S}\setminus\{\mu,\nu\}}^{{\cal P}\cup\bar{\cal P}
={\cal S}\setminus\{\mu,\nu\}}
\Bigg\{
\li_{\mu \nu}\,
\hat\rho^{\rm corr}_{\{\mu\}\cup{\cal P}}\,
\hat\rho^{\rm corr}_{\{\nu\}\cup\bar{\cal P}}
-
\tr_{\nu}\Bigg[
\li^S_{\mu \nu}
\sum_{{\cal Q}\subseteq\bar{\cal P}}
^{{\cal Q}\cup\bar{\cal Q}=\bar{\cal P}}
\hat\rho^{\rm corr}_{\{\mu\}\cup{\cal Q}}\,
\hat\rho^{\rm corr}_{\{\nu\}\cup\bar{\cal Q}}
\Bigg]
\hat\rho^{\rm corr}_{\{\nu\}\cup{\cal P}}
\Bigg\}
\nonumber \\
& &
+
\sum_{\mu \in {\cal S}} \li_\mu \hat\rho^{\rm corr}_{\cal S}
+ 
\frac{1}{Z}
\sum_{\kappa\notin{\cal S}} \sum_{\mu\in{\cal S}} 
\tr_{\kappa}\Bigg[
\sum_{{\cal P}\subseteq{\cal S}\setminus\{\mu\}}^{{\cal P}\cup\bar{\cal P}
={\cal S}\setminus\{\mu\}}
\li^S_{\mu \kappa}
\hat\rho^{\rm corr}_{\{\mu\}\cup{\cal P}}\,
\hat\rho^{\rm corr}_{\{\kappa\}\cup\bar{\cal P}}
\Bigg]
+\ord(Z^{-|\cal S|})
\,.
\eea
For $\ell=1$ and $\ell=2$, we recover equations 
(\ref{one-site-approx}) and (\ref{two-sites-approx}).
\end{widetext}

In contrast to the exact expression (\ref{general}), the approximated 
leading-order equations (\ref{general1}) form a closed set.
The exact time evolution (\ref{general}) of the $|\cal S|$-point 
correlator $\partial_t\hat\rho^{\rm corr}_{\cal S}$ also depends on the 
higher-order correlation term $\hat\rho^{\rm corr}_{{\cal S}\cup{\kappa}}$ 
involving $|{\cal S}|+1$ points. 
The approximated  expression (\ref{general1}), on the other hand, only 
contains correlators of the same or lower rank. 
This facilitates the iterative solution of the problem sketched in 
Section~\ref{hierarchyofcorr}.
First one solves the zeroth-order equation (\ref{one-site-approx}) 
for $\hat\rho_\mu^0$.
Inserting this result $\hat\rho_\mu^0$ into the first-order (in $1/Z$) 
equation (\ref{two-sites-approx}) for $\hat\rho_{\mu\nu}^{\rm corr}$, 
we obtain a first-order result for $\hat\rho_{\mu\nu}^{\rm corr}$. 
This first-order result for $\hat\rho_{\mu\nu}^{\rm corr}$ can then be 
inserted into the equation for $\hat\rho_{\mu\nu\lambda}^{\rm corr}$
which is of second order $1/Z^2$.
Furthermore, we may use the first-order result for 
$\hat\rho_{\mu\nu}^{\rm corr}$ in order to obtain a better approximation 
for the one-point density matrix $\hat\rho_\mu^1$ which is valid to first 
order in $1/Z$ and contains the quantum depletion etc. 
Repeating this iteration, we may successively ``climb up'' to higher and 
higher orders in $1/Z$. 



\begin{thebibliography}{99}
\bibitem{footnote-correlations}
Note that one should be careful what type of correlations are considered.
%
For example, the Mott state is often referred to as correlated because the 
many-body wave-function $\Psi(\f{r}_1,\f{r}_2,\dots)$ cannot be written as a 
symmetrized product $\Psi(\f{r}_1,\f{r}_2,\dots)
\neq{\cal S}[\psi_1(\f{r}_1)\psi_2(\f{r}_2)\dots].$
%
This corresponds to correlations between the particle positions such as 
$\f{r}_1$ and $\f{r}_2$, i.e., if one particle $\f{r}_1$ is at the lattice 
site $\mu$, the other particle $\f{r}_2$ is very likely not at the same site 
due to the strong repulsion $U$. 
%
In this paper, we do not consider this type of correlations 
(e.g., between $\f{r}_1$ and $\f{r}_2$) but instead the correlations 
between lattice sites $\mu$ and $\nu$.
\bibitem{footnote-lattice-structure}
Note that, in contrast to the $\ord(J)$-terms in 
Eqs.~(\ref{hp-temp-J^1}) and (\ref{hp-equil-J^1}), the existence of the 
$\ord(J^2)$-terms in Eqs.~(\ref{hp-equil-J^2}) and (\ref{hp-temp-J^2}) 
depends on the lattice structure.
%
In the case of a hyper-cubic lattice with rectangular tunneling to nearest 
neighbors only, for example, the sum over $\kappa$ in 
Eqs.~(\ref{hp-equil-J^2}) and (\ref{hp-temp-J^2}) contains at most two 
non-vanishing terms and thus the correlators in (\ref{hp-equil-J^2}) and 
(\ref{hp-temp-J^2}) are of order $1/Z^2$.
%
For a hyper-cubic lattice with additional diagonal tunneling, on the other 
hand, the $\kappa$-sum can involve $\ord(Z)$ non-zero contributions, which 
yields correlators (\ref{hp-equil-J^2}) and (\ref{hp-temp-J^2}) 
of order $1/Z$.
\end{thebibliography}
\end{document}